\documentclass{aa}

\usepackage{epsfig,graphicx,natbib}
\usepackage{longtable}
\usepackage{amssymb}
\usepackage{amsmath}
\usepackage{mathrsfs} 
\usepackage{color}
\usepackage{subcaption}
\usepackage{booktabs}
\usepackage{xspace}
\usepackage{float}
\usepackage{longtable}
\usepackage{breqn}
\usepackage{algpseudocode}
\usepackage{algorithm}
\usepackage{hyperref}

\usepackage{todonotes}
\usepackage{adjustbox}

\newcommand{\gapprox} {\, \lower3pt\hbox{$\sim$}\llap{\raise2pt\hbox{$>$}}\,}

\begin{document}

\title{Prospects of using closure traces directly for imaging in Very Long Baseline Interferometry}

\author{Hendrik M\"uller \inst{1,2}}

\institute{   Max-Planck-Institut für Radioastronomie, Auf dem Hügel 69, D-53121 Bonn (Endenich), Germany 
   \and 
   Jansky Fellow of National Radio Astronomy Observatory, 1011 Lopezville Rd, Socorro, NM 87801, USA \\ \email{hmuller@nrao.edu}
}

\date {Received  / Accepted}

\authorrunning{Hendrik M\"uller}

\abstract
{The reconstruction of the polarization of a source in radio interferometry is a challenging calibration problem since the reconstruction strongly depends on the gains and leakages that need to be inferred along with the image. This is particularly true for the Event Horizon Telescope (EHT) due to its small number of antennas, small signal-to-noise ratio and large gain corruptions.}
{To recover linear polarization, one either has to infer the leakages and gains together with the image structure, or rely completely on calibration independent closure quantities. While the first approach has been explored in Very Long Baseline Interferometry (VLBI) for a long time, the later one has been less studied for polarimetry.}
{Closure traces are a recently proposed concept of closure quantities \citep{Broderick2020b} that, in contrast to closure phases and closure amplitudes, are independent against both gains and leakages and carry the relevant information about the polarization of the source. Here we explore, how closure traces could be directly fitted to create an image and point out an imaging pipeline that succeeds in the direct imaging from closure traces.}
{Since closure traces have a number of inherent degeneracies, multiple local image modes that can fit the data are detected. Therefore, a multiobjective imaging technique is needed to correctly sample this multimodality.}
{Closure traces are not constraining enough for the current EHT configuration to recover an image directly, mainly due to the small number of antennas. For planned successors of the EHT however (with a significantly larger number of antennas), this option becomes feasible and performs competitive to the imaging with residual leakages.}


\maketitle

\section{Introduction}
In Very Long Baseline Interferometry (VLBI) multiple antennas are combined into a single array. The correlated signal of an antenna pair gives rise to the Fourier transform of the true sky brightness distribution (visibilities) with the spatial frequency defined by the baseline separating the two antennas projected on the sky plane, as described by the Cittert-Zernike theorem \citep{Thompson2017}. Imaging these data is an ill-posed inverse problem due to the sparsity of the coverage of measured Fourier coefficients (uv-coverage). Additionally, we have to deal with instrumental noise, and direction-independent, station-based antenna calibration effects at the imaging stage. That makes the imaging problem challenging particularly with the small phase stability when observing at high frequencies and in very sparse settings. 

VLBI observations in polarimetric light are particularly interesting for the study of Active Galactic Nuclei (AGN) jets since synchroton radiation is expected to be polarized. In fact, polarimetric studies of AGNs play a vital role in the understanding of the magnetic field in AGN \citep[e.g.][]{Homan2011, Gabuzda2021, Poetzl2021, Kramer2021, eht2021a, eht2021b, Ricci2022, Kramer2023}. However, polarimetric studies pose additional calibration challenges. Most importantly, we have to correct the measurements for leakage terms between orthogonal polarization feeds. These can be measured with observations of the calibrator source which appears ideally as a point source. However, when the calibrator source has poor parallactic coverage, or is resolved on its own as common in mm-VLBI experiments, the leakage estimation is more challenging.

Historically the imaging and calibration problem was solved by a hybrid inverse modelling approach \citep{Pearson1984}. In an alternating fashion, we image the data in total intensity, and self-calibrate the data. The imaging is done with the CLEAN algorithm \citep{Hogbom1974} or some of its successors \citep{Clark1980, Schwab1984, Bhatnagar2004, Cornwell2008, Rau2011, Offringa2017, Mueller2023a}. CLEAN is a greedy matching pursuit algorithm that models the image as a set of delta components in user-defined search windows. Self-calibration denotes the procedure of identifying the station-based gains that optimize the match between the model visibilities and the observed visibilities. In this sense, hybrid imaging could be understood as a two-step optimization procedure. Typically the data set is imaged and self-calibrated in total intensity, then the leakage terms are calibrated, and the imaging procedure is redone for all four Stokes parameters \citep{Cotton1993, Park2021, Marti2021}.

CLEAN is the de-facto standard in VLBI. However, CLEAN has well-known limitations \citep[e.g. see the discussions in][]{Mueller2023a, Pashchenko2023} that are most pronounced in sparse settings, e.g. see the performance comparisons in \citet{eht2019d, Arras2019, Mueller2022, Mueller2023a, Mueller2023d, ngehtchallenge}. 

Recently, forward modelling approaches have been proposed for sparse VLBI imaging in the framework of Regularized Maximum Likelihood (RML), see \citet{Honma2014, Akiyama2017a, Akiyama2017b, Chael2016, Chael2018, Chael2023}, compressive sensing \citep{Mueller2022, Mueller2023b}, Bayesian imaging \citep{Broderick2020, Arras2022, Tiede2022}, and multiobjective evolution \citep{Mueller2023c, Mus2023b, Mus2023c}. These reconstruction algorithms perform superior to CLEAN in terms of resolution, dynamic range and accuracy in an EHT-like setting. commented out general discussion about benefits of forward modelling

There are several strategies to adapt the self-calibration and leakage calibration in these forward modelling frameworks. These can be broadly divided into two opposite philosophies. First, we could add the unknown gains and leakage terms into the vector of unknowns and infer them during the imaging \citep[][]{JongKim2024} or alternating with the imaging \citep[e.g. with the strategy for time-dynamic and polarimetric reconstructions applied by][]{Mus2023b}. Second, we could fit directly to data products that are robust against gains and/or leakages, and ignore them in the procedure consequently. This has been realized by closure-only imaging in various imaging frameworks \citep{Chael2018, Mueller2022, Mueller2023c} that utilize closure phases and closure amplitudes. Moreover, calibration-robust polarization fractions were proposed for the reconstruction of the polarimetric signal for sparse mm-VLBI \citep{Johnson2017, eht2021a, Roelofs2023}. Note that \citet{Blackburn2020} has demonstrated that the direct fitting of self-calibration robust data products such as closures is equivalent to a Bayesian marginalization over completely unknown gains, e.g. with flat priors. Bayesian methods would only show an improvement if incorporating informative priors on these nuisance parameters. This mathematical equivalence is not necessarily true for RML approaches since the regularizers do not have in all cases a Bayesian interpretation, and overall the relative balancing between the data fidelity term and the regularization terms may be affected. Moreover, even for full Bayesian algorithms, this equivalence may not always been achieved in the numerical practice since marginalization over the gains and leakages may include integrating potentially high-dimensional multi-modal posterior distributions, which are only known approximately.

Closure traces are a novel concept of closure quantities introduced by \citet{Broderick2020b}. Closure traces are constructed from a quadrangle of antennas by a combination of the Stokes matrices. They are particularly independent against most types of data corruption, e.g. gains, antenna feed rotation, and leakage terms. Moreover, they contain closure phases and closure amplitudes as special cases in the unpolarized limit. A variant defined on triangles has been proposed by \citet{Joseph2022}, based on a full gauge theory for closure products \citep{Thyagarajan2022, Thyagarajan2022b}. It is natural to attempt to fit closure traces directly to infer the polarimetric features of a source. The use of closure traces for the polarimetric calibration may be imperative in all situations in which the parallactic angle coverage of the calibrator is insufficient for a proper calibration. A first pioneering work in this direction has been performed by \citet{Albentosa2023}, although in a model-fitting rather than an imaging framework. In this manuscript we study the prospects of constructing an imaging algorithm that, for the first time, fits directly to the closure traces rather than the visibilities.

Such an imaging application holds the specific advantage of being calibration-independent, thus avoiding the self-referencing, and necessarily local, hybrid imaging approach. However, closure traces are non-trivial complex quantities that are constructed out of the observables in an interferometric experiment (i.e. visibilities) and consequently the number of statistically independent closure traces is smaller than the number of independent visibilities. Roughly speaking, by constructing data products that are independent of the gains and leakages, we depend on less constraining data products, a fact that manifests by a number of degeneracies as listed by \citet{Broderick2020b}. To account for these degeneracies and increased sparsity during the imaging, we have to account for the missing information and non-convex optimization landscape by the use of more strict regularization information. One such approach, namely by multiobjective optimization \citep{Mueller2023c, Mus2023c} is presented in this manuscript. 


For this work we focus on two cases. First we study VLBI, specifically the EHT and its planned successors. This data regime stands out as a particularly interesting area of application of calibration-independent imaging since the high frequency, global baselines, and sparsity of the array render the initial calibration (especially with respect to phases) very challenging \citep{Janssen2022}, and the imaging/self-calibration may be at risk to fall into local extrema and method-specific biases. 

Second, we study the application of closure trace imaging in denser radio interferometers, represented exemplary by ALMA where we reproduce the experiment performed by \citet{Albentosa2023}. While the EHTC has contributed a wide variety of imaging, and calibration tools (including the forward modelling techniques and closure quantities studied in this manuscript), and continues to be the target of active and ongoing developments, a transfer of these ideas to general interferometric imaging has been done only on a few occasions \citep[e.g. recently][]{Albentosa2023, Fuentes2023, Lu2023, Kramer2023, Mueller2023d}. 
By testing closure trace imaging with a real ALMA observation, we also aim to contribute towards the transfer of knowledge and methodologies gained within the EHTC for general interferometric experiments.

In pursue of these efforts towards imaging directly from the closure traces, both RML approaches \citep{Honma2014, Chael2016, Akiyama2017a, Akiyama2017b, Chael2018, Mueller2022, Mueller2023b, Mueller2023c, Chael2023, Mus2023c} as well as Bayesian approaches \citep{Broderick2020, Pesce2021, Arras2022, Tiede2022} have been developed and improved over the years, existing now in a matured software state. This is a formidable variety of options to tackle the challenges arising from directly fitting closure traces. For this study we are using the recently proposed multi-objective evolution strategies \citep{Mueller2023c, Mus2023b, Mus2023c} since they are utilizing the full set of regularization terms frequently used for RML and are proven to effectively sample potentially multimodal distributions. This last property is expected to be crucial for the fitting of closure traces, since the reduced number of statistically independent observables, the degeneracies inherent to closure traces, and finally the non-convex nature of the forward operator, leave local search techniques at risk at falling into local minima, opposed to multiobjective imaging algorithms which navigate the objective landscape globally. However, while more robust against non-convex problem settings, multiobjective minimization strategies depending on decomposition may also be trapped in local minima even when the full Pareto front is explored, since the scalarized subproblems may be non-convex as well.

\section{Theory}

\subsection{VLBI}
For VLBI multiple antennas, deployed at up to global baselines, observe the same source and act as an interferometer. The correlated signal of each antenna pair is proportional to the Fourier transform of the on-sky image. The associated Fourier frequencies are the baselines separating the antennas projected to the sky plane. The forward problem for total intensity imaging is thus described by the van Cittert-Zernike theorem:
\begin{align}
    \mathcal{V}(u,v) = \mathcal{F} I(l,m).
\end{align}
Here $\mathcal{V}$ denotes the observables (visibilities), $u,v$ are harmonic coordinates in the units of wavelengths describing the baseline coordinates, $\mathcal{F}$ is the Fourier transform operator, $I$ the true image, and $l,m$ direction-cosines describing on-sky coordinates relative to the phase center.

The difficulty in VLBI imaging arises from multiple complications. First, each antenna pair at each integration time determines only a single Fourier component. While the baseline relative to the on-sky source is varying in time due to Earth rotation, the limited number of antennas in an array limits the number of observed visibilities, resulting in a sparse representation of the Fourier domain. The coverage of measured Fourier coefficients is commonly referred to as uv-coverage. Second, we deal with antenna-based corruptions arising from a variety of effects such as pointing, focus, receiver sensitivity, and most importantly local weather conditions that may be prone to vary on rapid time-scales. These effects are typically factored into station-based gains $g_i$, complex numbers that are allowed to vary across time. Finally, the observation is corrupted by thermal noise $N$ which is typically modelled by a Gaussian distribution. The observed visibility at an antenna pair $i,j$ at a time $t$ is thus:
\begin{align}
V_{i,j,t} = g_i g_j^* \mathcal{V}(u_{i,j,t}, v_{i,j,t}) + N_{i,j,t}.
\end{align}

\subsection{Polarimetry}
For polarimetric observations, ideally each antenna observes the source with two polarization channels forming an orthogonal basis (either orthogonal linear, or left- and right-handed circular polarization feeds). However, for the EHT which is one focus point of this analysis, the JCMT observes in only one channel \citep[e.g. compare][]{eht2021a}. 

The correlation of an antenna pair now yields four independent correlation products: $V^{RR}_{i,j,t}, V^{RL}_{i,j,t}, V^{LR}_{i,j,t}, V^{LL}_{i,j,t}$ where $R$ denotes right-handed polarization and $L$ denotes left-handed polarization. These correlation products are refactored into four Stokes parameters: $I, Q, U, V$, where $I$ describes the total intensity, $Q,U$ linear polarization and $V$ circular polarization.
The Stokes visibilities are, similarly to the total intensity case, related to the polarized image structures by the van-Cittert/Zernike theorem:
\begin{align}
    &\mathcal{V}_I =  \mathcal{F}I, \\
    &\mathcal{V}_Q =  \mathcal{F}Q, \\
    &\mathcal{V}_U =  \mathcal{F}U, \\
    &\mathcal{V}_V =  \mathcal{F}V.
\end{align}
The Stokes visibilities are commonly represented by the 2x2 visibility matrix:
\begin{align}
\mathbf{\mathcal{V}}_{ij} = \left( \begin{array}{cc}
\mathcal{V}_I^{ij}+\mathcal{V}_V^{ij} & \mathcal{V}_Q^{ij}+i \mathcal{V}_V^{ij} \\
\mathcal{V}_Q^{ij}+i \mathcal{V}_V^{ij} & \mathcal{V}_I^{ij}-\mathcal{V}_V^{ij}
\end{array} \right).
\end{align}
A graphical representation of the polarization state is the Poincare-sphere, see Fig. \ref{fig:poincare-sphere}. The Poincare-sphere depicts the linear polarization state with a longitude, and the circular polarization state by latitude on a sphere.

The Stokes parameters satisfy the inequality:
\begin{align}
    I^2 \geq Q^2+U^2+V^2. \label{eq: pol_ieq}
\end{align}
This motivates the representation of the linearly polarized structure by the polarization fraction $m$ and mixing angle $\chi$:
\begin{align}
    P = Q + \mathrm{i} U = I \cdot m \cdot \exp(\mathrm{i} \chi). \label{eq: mchi}
\end{align}

Polarimetric reconstructions now deal with the problem of recovering the image in all four Stokes parameters $I,Q,U,V$ from the Stokes visibilities (the observables). However, similar as in the case for total intensity, the image reconstruction is severely challenged by the uv-coverage. Moreover, we have to deal with gain and leakage corruptions as well. In the convention of visibility matrices, the gain corruptions are represented by a Jones-matrix as well $\mathbf{J}^i_{gain}, \mathbf{J}^i_{rotation}, \mathbf{J}^i_{leakage}$.
The observed visibilities are related to the unperturbed visibilities by \citep{Hamaker1996}:
\begin{align}
 \mathbf{V}_{i,j} = \mathbf{J}^i \mathcal{V}_{i,j} \left(\mathbf{J}^j \right)^\dag+\mathbf{N}_{i,j},
\end{align}
where $\mathbf{J}^i = \mathbf{J}^i_{gain} \mathbf{J}^i_{leakage} \mathbf{J}^i_{rotation} $ denotes the full corruption matrix of antenna $i$. The leakage corruptions are of particular importance, since they introduce off-diagonal terms.

\subsection{Closure Quantities} \label{sec: closure_quantities}
Closure quantities are quantities that are derived from the observed visibilities. They are constructed in a way that they are independent of gain and/or leakage corruptions. Closure phases and closure amplitudes were used for a long time for total intensity imaging in VLBI. Closure phases are the phase derived over a triangle of antennnas:
\begin{align}
   \Psi_{ijk} = arg \left( V_{ij} V_{jk} V_{ki} \right)\,.
\end{align}
Closure amplitudes are defined over a quadrilateral of antennas:
\begin{align}
    A_{ijkl} = \frac{|V_{ij}| |V_{kl}|}{|V_{ik}| |V_{jl}|}\,.
\end{align}
Closure phases and closure amplitudes are independent of the gains, and thus have been used for closure-only imaging in the past \citep{Chael2018, Blackburn2020, Arras2022, Mueller2022, Mueller2023c}.

Closure traces could be understood as the polarimetric counter-part of closure amplitudes \citep{Broderick2020b}. They are defined over a quadrilateral of antennas by:
\begin{align}
    Tr_{i,j,k,l} = \frac{1}{2} \mathrm{trace} \left( \mathbf{V}_{i,j} \mathbf{V}^{-1}_{k,j} \mathbf{V}_{k,l} \mathbf{V}^{-1}_{i,l} \right).
\end{align}
It can be shown now that $Tr_{i,j,k,l}$ is independent of $\mathbf{J}^{i},\mathbf{J}^{j},\mathbf{J}^{k},\mathbf{J}^{l}$, i.e. independent of gains and leakage terms \citep{Broderick2020b}. On one quadrilateral with four different antennas $i \neq j \neq k \neq l$, there are 24 permutations of $i,j,k,l$. However, only six out of these combinations are independent. These are \citep[for a proof and more details, see][]{Broderick2020b}:
\begin{align}
    CT_{i,j,k,l} = \left[ Tr_{i,j,k,l}, Tr_{i,j,l,k}, Tr_{i,k,j,l}, Tr_{i,k,l,j}, Tr_{i,l,j,k}, Tr_{i,l,k,j}  \right]. \label{eq: CT}
\end{align}

\begin{figure}
    \centering
    \includegraphics[width=0.5\textwidth]{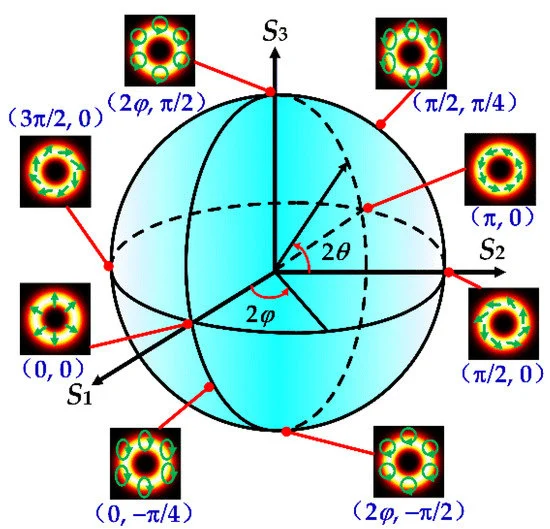}
    \caption{Illustration of the Poincare-sphere, reprinted from \citet{Ma2022} under CC.BY 4.0.}
    \label{fig:poincare-sphere}
\end{figure}

While closure phases and closure amplitudes have been directly fitted for total intensity imaging, closure traces have not been used as a calibration-independent data product that gets directly fitted by an imaging algorithm. However, there are first attempts to fit closure traces by model-fitting \citep{Albentosa2023}, and the analysis of closure traces have played a crucial role in the pre-imaging analysis of EHT data sets \citep{eht2021a, eht2023}. This may be caused by a number of degeneracies inherent to closure traces. As closure phases and closure amplitudes, they do not contain the absolute position, and the total flux of the image. However, the information regarding the relative information of structural features, their relative flux and the relative polarization signal remains preserved \citep[as recently demonstrated by][]{Albentosa2023}. Particularly, closure traces are independent against any rotation on the Poincare-sphere \citep{Broderick2020b}. As a special case, that mean that absolute EVPA information is lost (rotations along the equator of the sphere). However, rotations on the Poincare-sphere also include rotations in latitude, mixing up linear polarization and circular polarization, compare Fig. \ref{fig:poincare-sphere}. 

Closure traces are defined on quadrilaterals. They will be the main focus of this manuscript due to their widespread use within the EHT collaboration. However, an alternative concept defined on closure triangles has been developed recently as well \citep{Thyagarajan2022, Joseph2022}. This concept builds invariants by triangles to a central station, and processing the Minkowski product of the associated four vectors. 

\section{Imaging}

\subsection{Overview}
Imaging has been historically performed by the CLEAN algorithm and its variants \citep{Hogbom1974, Clark1980, Schwab1984, Bhatnagar2004, Cornwell2008, Rau2011, Offringa2017, Mueller2023a}. Recently forward modelling techniques have been proposed for VLBI as an alternative, especially for the EHT. These can be grouped into various categories: Bayesian reconstruction algorithms \citep{Broderick2020, Tiede2022, Arras2022}, RML approaches \citep{Akiyama2017a, Akiyama2017b, Chael2018}, compressive sensing \citep{Mueller2022}, multiobjective imaging \citep{Mueller2023c}, and swarm intelligence \citep{Mus2023c}. Most of the aforementioned algorithms have been extended successfully to the polarimetric domain \citep{Chael2016, Broderick2020, Tiede2022, Mueller2023b, Mus2023b, Mus2023c}. While the imaging algorithms are implemented in a variety of softwares, and differ significantly in the way the image is represented and regularization is perceived, all the forward modelling techniques have a relative modularity in common making it easy to extend the imaging algorithms to new data terms. 

In this manuscript, we study the direct imaging from closure traces from the perspective of RML methods, although Bayesian frameworks may be equally well qualified as long as they support multimodal posterior exploration \citep[in contrast to e.g. resolve][]{Arras2022}. This is the first direct imaging attempt from the closure traces. To study how restricting closure traces are, and what imaging strategies may be needed, we study two competitive, alternative approaches.

First we utilize the numerically expensive multiobjective imaging algorithm MOEA/D recently proposed by \citet{Mueller2023c} and extended to polarimetric, and time-dynamic reconstructions in \citet{Mus2023b}. MOEA/D uses the full set of common RML regularization terms \citep[as they were applied in][]{eht2021a}, and searches for the front of Pareto optimal solutions among them. Among all RML approaches, MOEA/D may be most suited to study calibration independent imaging, since instead of a single (maximum-a-posteori) solution, a list of locally optimal solutions is found. In this way, MOEA/D may identify degeneracies, possibly multimodality and non-convexity. MO-PSO as proposed by \citet{Mus2023c} is a variant of MOEA/D based on particle swarm optimization that does not explore the full Pareto-front, but solves several limitations of MOEA/D and improves over MOEA/D in accuracy. 

Second, in contrast to MOEA/D, we utilize a lightweight alternative: DoG-HiT \citep{Mueller2022}. DoG-HiT represents the image by wavelets that are fitted to the uv-coverage \citep{Mueller2022, Mueller2023a} and has been extended to polarimetry and dynamics \citep{Mueller2023b}. All RML regularization terms are replaced by a single, wavelet-based, regularizer. In this way, DoG-HiT, contrasting MOEA/D, represents a simpler, automated alternative with weaker, and data-driven regularization information.

With studying the closure-trace imaging from these perspectives, we cover the spectrum of RML methods along relatively weak prior assumptions (DoG-HiT) and relatively strong prior information (MOEA/D), allowing to identify the future prospects of imaging directly with the closure traces. We summarize in Appendix \ref{app:imaging} the essential outline of the imaging algorithms MOEA/D, MO-PSO, and DoG-HiT. For more details we refer to the respective references therein, especially \citet{Mueller2022} for DoG-HiT, \citet{Mueller2023c, Mus2023b} for MOEA/D and \citet{Mus2023c} for MO-PSO.

\subsection{Imaging Procedure}\label{sec: imaging}

Recall from Sec. \ref{sec: closure_quantities}, and particularly the discussions in \citet{Broderick2020b}, that closure traces have a variety of internal degeneracies. The most problematic one for a direct imaging from the closure traces is their property of being independent against any rotation in the Poincare-sphere. While closure traces are independent against any global rotation, their relative position across the image remains kept. In the case of neglegible circular polarization, this rotation invariance translates into a degree of freedom in the overall EVPA orientation that could be calibrated effectively by accompanying observations on calibrator sources, e.g. see the strategy applied in \citet{Albentosa2023}. This degeneracy (lost global EVPA orientation) is described by the rotations around the equator of the Poincare-sphere. However, if significant circular polarization is detectable in the source of interest, rotation invariance around the latitude could easily mix up linear polarization and circular polarization. We therefore restrict ourselves in this work to situations with negligibly small circular polarization at first instance. Among the rotation invariance, additional, potentially more complex degeneracies may be expected that need to be taken into account during the imaging. 

Compared to using the full set of visibilities, any kind of closure quantities represents just a subset of observables. This contributes to the sparsity of the observation, and ultimately worsens the situation of an only weakly constraining data set. In the case of closure traces it can be shown that they are a statistically complete set of all gain and d-term independent quantities. It has been demonstrated in the past that imaging from closure products only, at least in total intensity, is still feasible when powerful regularization frameworks are used. This has been originally demonstrated for RML imaging in \citet{Chael2018}, and was subsequently utilized in a wide range of imaging approaches in the context of the EHT including Bayesian reconstructions \citep{Arras2022}, compressive sensing \citep{Mueller2022}, and multiobjective imaging \citep{Mueller2023c}.

Finally, non-convexity may be a potential issue for the imaging. Strict convexity of the objective is a sufficient condition to prove uniquness of an optimization, and a single-modal posterior. The opposite is not necessarily true, i.e. a non-convex optimzation problem does not necessarily need to be multimodal, but is prone to be so in practice. The $\chi^2$-metric to the closure traces is not strictly conex. This is easily proven by the existence of a flat plateau along the rotation-invariance around the Poincare-sphere, as discussed above. However, we can also not analytically prove that there is a definitely concave part of the objective.

In any case, these three limitations (inherent degeneracies, a smaller number of independent observables, and a potentially non-convex optimization problem) pose serious issues for local search methods, that in consequence may be trapped in local extrema rather than searching for the global optimum. Multiobjective optimization, and particularly MOEA/D \citep{Mueller2023c}, are a good alternative. Due to their randomized search techniques they navigate the objective landscape more globally. Particularly, the various clusters of Pareto optimal solutions identified by MOEA/D both in total intensity, as also in polarimetric reconstructions \citep{Mus2023b} are interpreted as lists of locally optimal modes. Therefore, we are using MOEA/D in the following.

The set-up of MOEA/D for the reconstruction with respect to closure traces consists of solving a multi-objective problem with the following objective functionals \citep[inspired by the parametrization proposed in][]{Mus2023b}:
\begin{align}
    & f_1:=\alpha \chi^2_{CT}+\beta R_{ms}, \label{eq: pol1}\\
    & f_2:=\alpha \chi^2_{CT}+\gamma R_{hw},\label{eq: pol2}\\
    & f_3:=\alpha \chi^2_{CT}+\delta R_{ptv},\label{eq: pol3}\\
    & f_4:=\alpha \chi^2_{CT}. \label{eq: pol4}
\end{align}
Here we used the $\chi^2$-statistics to the array of six independent closure traces $CT$, see Eq. \eqref{eq: CT}, as data term. The regularization terms are polarimetric variants of the entropy $R_{ms}, R_{hw}$ and the polarimetric counterpart of the total variation regularizer $R_{ptv}$. The exact regularization terms can be retrieved by Appendix \ref{app:regularizers}.


The specific challenges posed by directly fitting the closure traces made a couple of additional updates to the imaging pipeline of MOEA/D necessary that are detailed in Appendix \ref{app:moead_updates}.

\section{Synthetic Data Tests}
We test the potential for closure trace imaging in a variety of settings and along multiple axes with targeted tests. In fact, we study whether the imaging works in general and is competitive to classical approaches, and how that assessment changes with the sampling of uv-domain, and methodology used.

\subsection{Synthetic Data}
\begin{figure*}
    \centering
    \includegraphics[width=\textwidth]{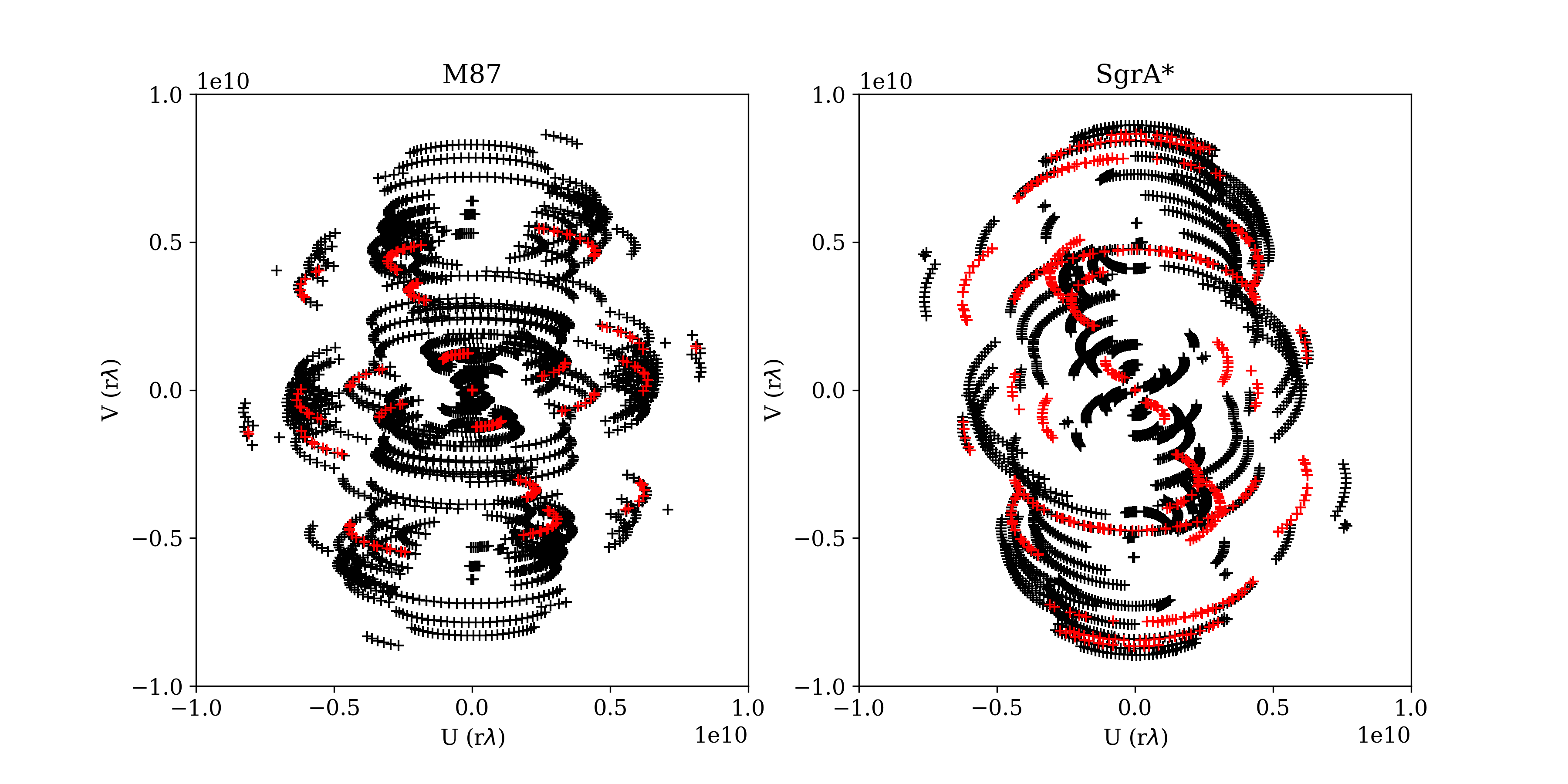}
    \caption{Illustration of the uv-coverage on M87 (left panel), and SgrA* (right panel) with the EHT array of 2017 (red points) and the proposed EHT+ngEHT configuration \citep[black points, ][]{ngehtchallenge}.}
    \label{fig:uvc}
\end{figure*}

To this end, we perform the synthetic data tests with the configuration of the EHT in 2017, and with a proposed configuration of the ngEHT taken from the ngEHT Analysis challenge \citep{ngehtchallenge}. The latter one has in particular been selected since the corresponding data sets and data properties have been regularly been used to verify and validate novel imaging algorithms. The array consists of all the current EHT antennas and ten additional stations taken from the list of \citet{Raymond2021}. We compare the corresponding uv-coverages in Fig. \ref{fig:uvc}.

We test three synthetic ground truth models in this manuscript. One model is a geometric model that has similar properties as the EHT observations of M87 in polarization \citep{eht2021a}. It consists of a ring with a diameter of $42\,\mu\mathrm{as}$, and a compact flux of $0.6\,\mathrm{Jy}$. The EVPA pattern is a radial pattern twisted by $45$ degrees. The model is equivalent with the geometric ring model with a vertical magnetic field that was tested in \citet{Mus2023b}.

As an additional test source, we use the result of a GRMHD simulation of SgrA* that is commonly available in the ehtim software package\footnote{\url{https://github.com/achael/eht-imaging}}. The model mimics the typical crescent-like appearance of SgrA* in EHT observations, with a polarimetric pattern that differs around the ring, consequently covering a different test case than the first model.

Finally, we test a non-ring mode. To this end, we developed a synthetic data set inspired by the observation of 3C279 with the EHT \citep{Kim2020} which essentially resembles a double source with reshaped elongated gaussians. The ground truth image is constructed out of six gaussian components (C0-0, C0-1, C0-2, C1-0, C1-1, C1-2) with the amplitudes and relative positions that were reported from the geomtric model fitting results shown in \citet{Kim2020} for the 05. April. We separate the two clusters and add linear polarization synthetically with an EVPA pattern roughly aligned with the expected jet axis in 3C279 \citep{Kim2020}.

The synthetic observations have been performed with the EHT and the EHT+ngEHT configuration respectively. To this end, we load the true observations done with the EHT on April 11. 2017 \citep{eht2019a} from the public release page of the EHT \footnote{\url{https://datacommons.cyverse.org/browse/iplant/home/shared/commons_repo/curated/EHTC_M87pol2017_Nov2023}},  and the synthetic ground truth data set used in the ngEHT analysis challenge \footnote{retrieved from \url{https://challenge.ngeht.org/challenge1/}} for M87. For the synthetic SgrA* reconstructions we load the respective observation files on SgrA* from the same webpages \footnote{real data can be retrieved from \url{https://datacommons.cyverse.org/browse/iplant/home/shared/commons_repo/curated/EHTC_FirstSgrAPol_Mar2024}}. The synthetic observations have been performed by the \texttt{ehtim} package. This routine basically reads the uv-coordinates, and respective thermal noise level, from an existing observation file, evaluates the Fourier transform of the ground truth model at these spatial frequencies and adds thermal noise. This is a rather simplistic forward model, and we would like to note the existence of more complex simulation tools that were developed within the EHT collaboration \citep[e.g.][]{Roelofs2020}, that mainly model tropospheric effects with more richness of detail. In the case of SgrA*, we ignored the effects of the scattering for now, e.g. we did not apply a strategy of inflating the noise-levels as was done by the EHT \citep{eht2022c}. Since the synthetic observations mimic the systematic noise-levels of previously reported real and synthetic data sets, we refer to \citet{eht2019d, eht2022c, ngehtchallenge} for more details on the data properties. For the original EHT+ngEHT data sets simulated by \citet{ngehtchallenge}, we assumed a bandwidth of $8\,\mathrm{GHz}$ and a receiver temperature of $60\,\mathrm{K}$ with an aperture efficiency of $0.67$ at 230 GHz. The opacity values were extracted as the median of the estimates reported by \citet{Raymond2021} at an elevation of 30 degrees \citep{ngehtchallenge}.

We add gain corruptions and d-term corruptions to the observations. The d-terms have been perturbed by a mean of $5\%$ on all baselines, not taking into account typical d-term signatures for the EHT (e.g. the fact that d-terms for ALMA are typically only imaginary). In any case, these corruptions do not affect the closure traces at all, and the imaging remains robust against gain and d-term corruptions. We scan-average the data and rescale the inter-baseline fluxes to the correct total flux before forming closure triangles and closure quadrangles. A systematic noise budget of $2\%$ has been added to the data on all baselines before reconstruction.

\subsection{Reconstruction Results}
\begin{figure*}
    \centering
    \includegraphics[width=\textwidth]{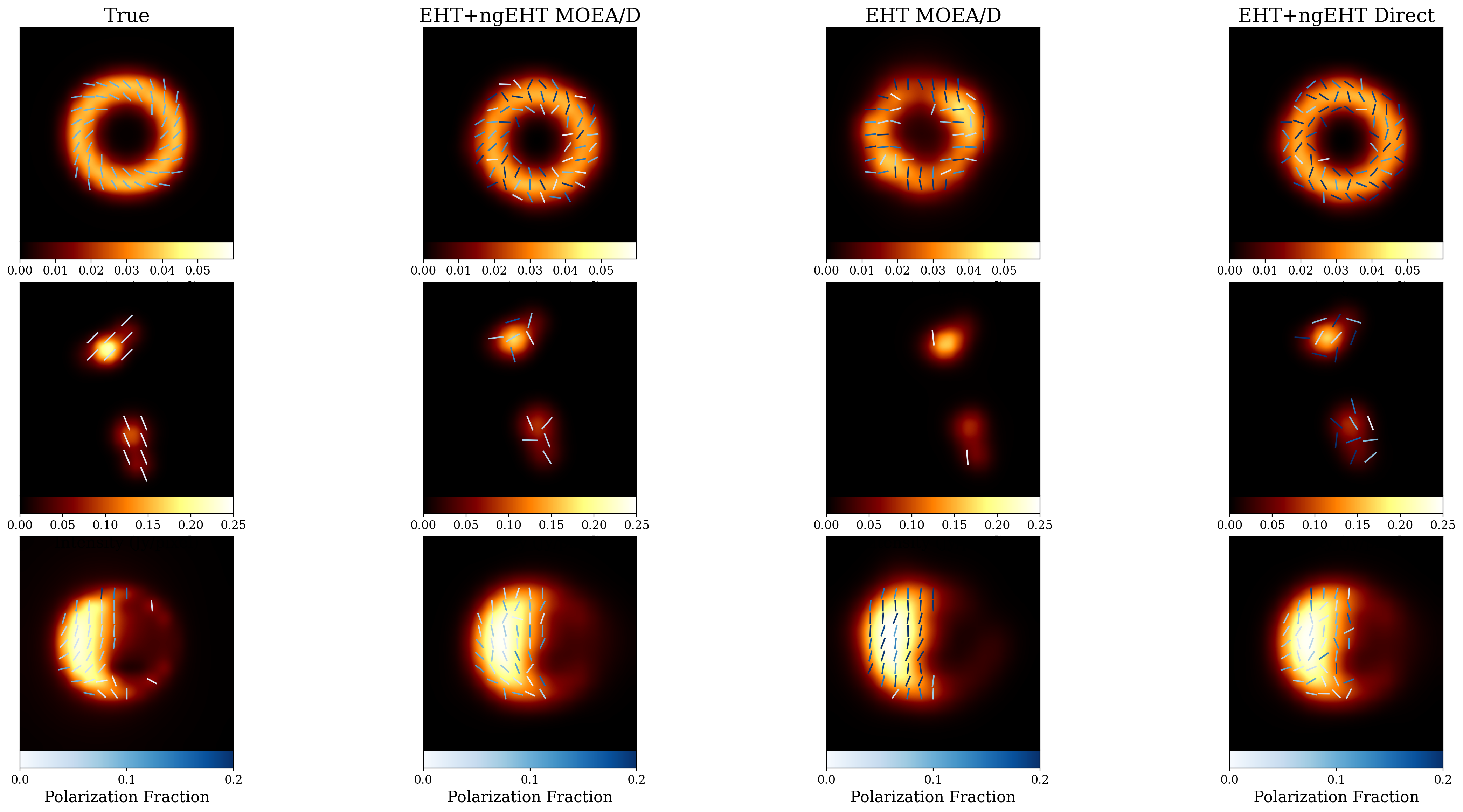}
    \caption{Comparisons of three different source structures (ring inspired by  \citet{eht2021a}, a double source mimcking \citet{Kim2020}, and a GRMHD model of SgrA*) and their corresponding reconstructions. The first column shows the ground truth, the second and third column the reconstructions with closure trace imaging with MOEA/D with two different array configurations. The fourth column shows a direct reconstruction with DoG-HiT with $5\%$ of residual (not corrected) leakages as a comparison.}
    \label{fig: comp}
\end{figure*}

We show the reconstruction results in Fig. \ref{fig: comp}. We compare the reconstructions with MOEA/D only fitting to closure traces for the EHT, the EHT+ngEHT configuration, and with a reconstruction that directly fits the (non-calibrated for leakages) Stokes visibilities. The latter reconstruction just assumes that the emission in total intensity, and the polarized emission share similar spatial scales and support (e.g. have a similar size of the compact emission). No explicit regularization terms on the polarized emission, such as a polarized entropy or smoothness assumptions, are used. Thus, it represents a rather weak prior information that does not process the closure traces. We did not fit d-terms, such that the reconstruction shows the effects of a missing d-term calibration in comparison to a d-term independent reconstruction.

\begin{figure*}
    \centering
    \includegraphics[width=\textwidth]{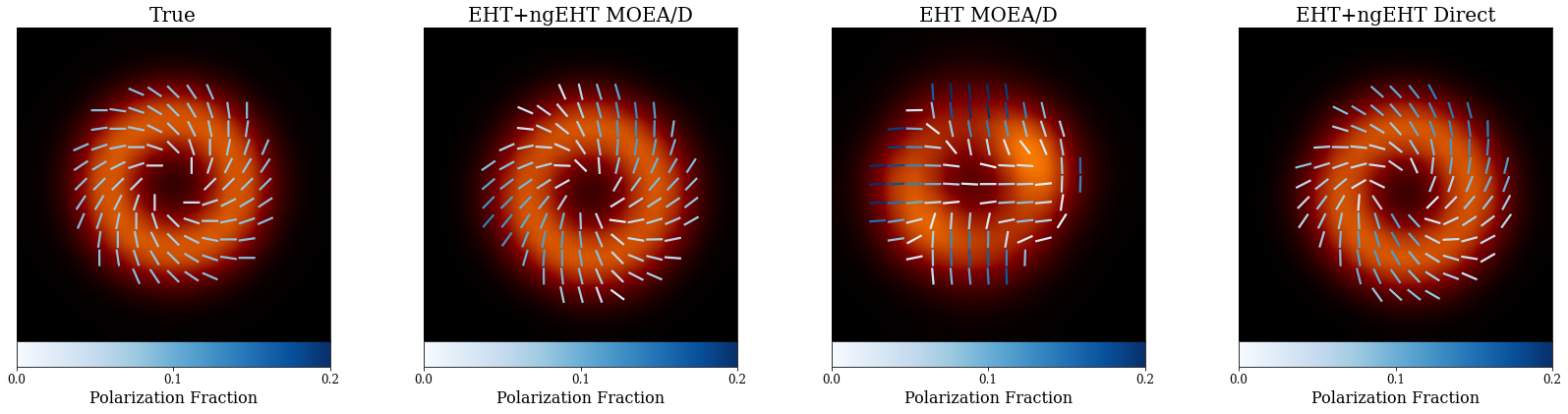}
    \caption{Panels from the first line in Fig. \ref{fig: comp}, but with a blurring of $20\,\mu\mathrm{as}$ applied.}
    \label{fig: comp_blurred}
\end{figure*}

The reconstruction with the EHT+ngEHT configuration is rather successful in both cases. The reconstructions look more scrambled in the reconstruction than in the ground truth. In Fig. \ref{fig: comp_blurred} we show the reconstructions blurred with the nominal beam resolution ($20\,\mu\mathrm{as}$). This comparison at lower resolution shows that the scrambled, more noisy reconstruction seems to be an effect of the reconstruction of super-resolved structures. 

While the EHT+ngEHT configuration allows for well reconstruction results, the EHT observations with a much sparser uv-coverage are less successful. This is understandable giving the much smaller number of antennas. The EHT observations did use seven antennas. Additionally, one antenna (JCMT) needed to be flagged for the closure trace analysis, since only one polarization channel is available. Moreover, two antenna pairs (ALMA/APEX, and JCMT/SMA) are essentially at the same position. Hence, the number of quadrilaterals that resolve structure is small for the EHT configuration. In fact, the EHT reported only one quadrilateral with sufficient signal to noise ration to draw conclusions \citep{eht2021a, eht2023}.

Since the imaging is independent of any kind of d-terms, this challenging analysis step can be omitted. However, at a backside, the number of independent observables drops, making the imaging more challenging. The imaging directly with closure traces is only a valid methodology, if the loss in sensitivity is smaller than the loss in accuracy introduced by the d-terms. Comparing the second and last column in Fig. \ref{fig: comp} and Fig. \ref{fig: comp_blurred}, this seems to be the case qualitatively. We will present a closer look at this property in Sec. \ref{sec: dterms}.

\subsection{Pareto Fronts} \label{sec: pareto}

Now let us have a closer look at the Pareto fronts. The Pareto fronts are presented in Appendix \ref{app:pareto_fronts}. We show the Pareto fronts in the figures Fig. \ref{fig: model1}, Fig. \ref{fig: model1_blurred}, and Fig. \ref{fig: model1_eht_blurred} for the ring model. MOEA/D finds a number of clusters representing locally optimal modes. The clusters 0, 1, and 2 are dominated by the polarimetric total variation and lying at the edge of the parameter space, hence do not describe the ground truth image properly. However, there are two clusters with well scoring with respect to the closure traces. Interestingly, they do describe two different radial EVPA patterns, describing two different local minima that a local search technique may fall into. MOEA/D succeeds in sampling both, and thus describes the range of possible image structures. 

In practical situations, we do not know the ground truth image. \citet{Mueller2023c, Mus2023b} developed two different approaches to select the objectively best image. A selection by the accumulation point of the Pareto front, or by the cluster that is closest to the utopian/ideal proved successful in practice, both for total intensity as well as for polarimetric reconstructions. Especially the distance to the ideal point was then used in \citet{Mus2023c} as the objective criterion for particle swarm optimization. We mark the solution that is closest to the ideal by a red box in Fig. \ref{fig: model1}, \ref{fig: model1_blurred}, \ref{fig: model1_eht_blurred}. The selection criterion selects the best image, the same is true for the double source and the GRMHD example as well, see Fig. \ref{fig: double} and Fig. \ref{fig: sgra}.

In the case of closure traces, there is furthermore an alternative possible concept for the selection. Since two different concepts for closure products exist, closure traces defined on quadrilaterals \citep{Broderick2020b} and closure invariants defined on closed triangles by gauge theory \citep{Thyagarajan2022, Joseph2022}, one could try to retrieve the best cluster by the scoring in the triangle based closure invariants. Although closure traces are a complete set of closure products, the closure triangle version defined by \citet{Thyagarajan2022, Joseph2022} may package this information in a different way. A different packaging of in principle equivalent information in the data fidelity term could impact the reconstruction in a RML framework, at least due to the locality of the optimization methods. Furthermore, the noise distribution on the closure triangles and closure quadrangles is only known approximately, and differs between the triangle relations and quadrangle relations. Therefore, a cross-validation may find the best reconstruction. We mark the cluster selected by the best fit to closure triangles with a dashed blue box. However, this selection fails. This may be explained by the fact that both closure invariants form a full set of invariant properties, and encode the same amount of information consequently, making a cross-verification not the correct tool to select the best cluster.

\subsection{MO-PSO}
The selection of the best cluster discussed in Sec. \ref{sec: pareto} suggests that the selection of the most natural cluster by the smallest distance to the ideal point works. This distance was proposed as the objective functional for MO-PSO \citep{Mus2023c}. MO-PSO presents some advantages over MOEA/D. Since the evolutionary algorithm is only used to explore the weights space, but not to recover the actual image from the visibility data, it is orders of magnitude faster than MOEA/D. MO-PSO does not recover the full set of local extrema as MOEA/D (which was identified crucial for this application), but it still searches the objective landscape globally and converges to the same preferred cluster as MOEA/D (since the selection criterion is used as an objective functional). Based on the analysis presented in the previous subsection, MO-PSO is also applicable for the problem of recovering polarimetric images directly from the closure traces. 

\begin{figure*}
    \centering
    \includegraphics[width=\textwidth]{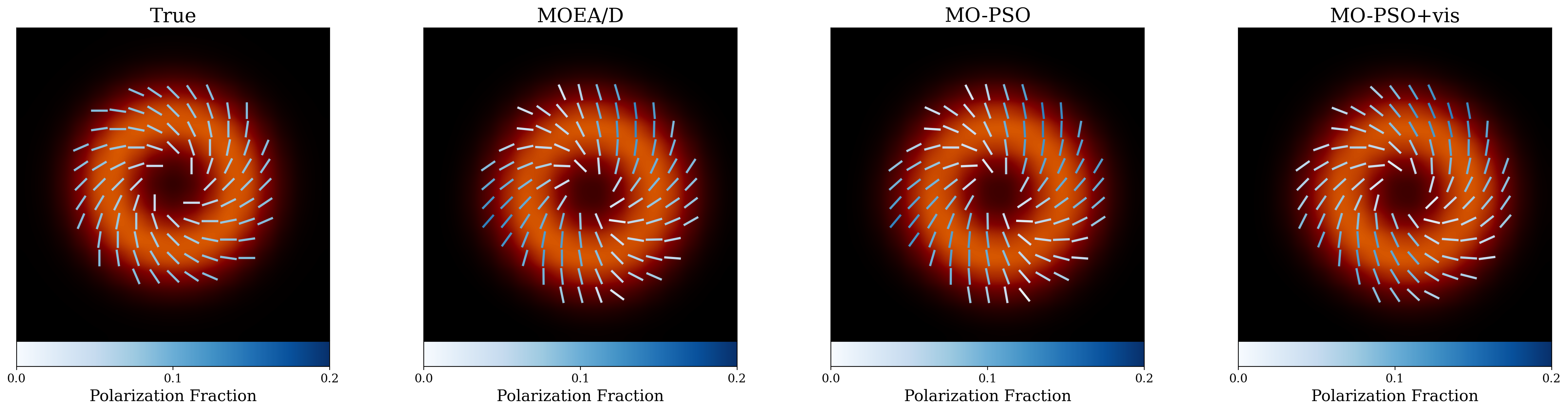}
    \caption{Comparison of the ground truth image compared to various realizations of directly fitting closure traces for the EHT+ngEHT configuration. Second column: reconstruction by MOEA/D only fitting closure traces. Third column; reconstruction by MO-PSO only fitting closure traces. Fourth column: reconstruction by MO-PSO fitting closure traces and visibilites.}
    \label{fig: pso_ngeht}
\end{figure*}

\begin{figure*}
    \centering
    \includegraphics[width=\textwidth]{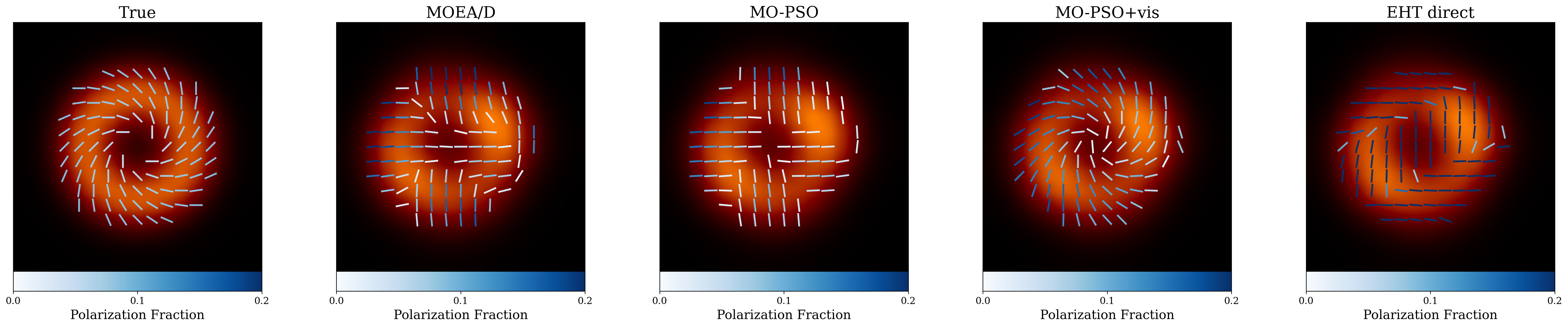}
    \caption{The same as Fig. \ref{fig: pso_ngeht}, but for the EHT configuration. Additionally we show in the fifth column a direct reconstruction from the visibilities when leakages at a level of $5\%$ are not corrected.}
    \label{fig: pso_eht}
\end{figure*}

We show in Fig. \ref{fig: pso_ngeht} the reconstruction result of MO-PSO for the EHT+ngEHT configuration, and in Fig. \ref{fig: pso_eht} the performance for the EHT array alone. The reconstructions with MO-PSO and MOEA/D do not differ much, while the latter one is performed much faster. In conclusion, since multiple clusters have been identified with MOEA/D, global search techniques should be preferred over local search techniques. However, the success of MO-PSO for a denser array configuration indicates that the full Pareto front does not need to be sampled, rather the most natural solution is preferred.

\subsection{Effect of leakage inaccuracies} \label{sec: dterms}
\begin{figure*}
    \centering
    \includegraphics[width=\textwidth]{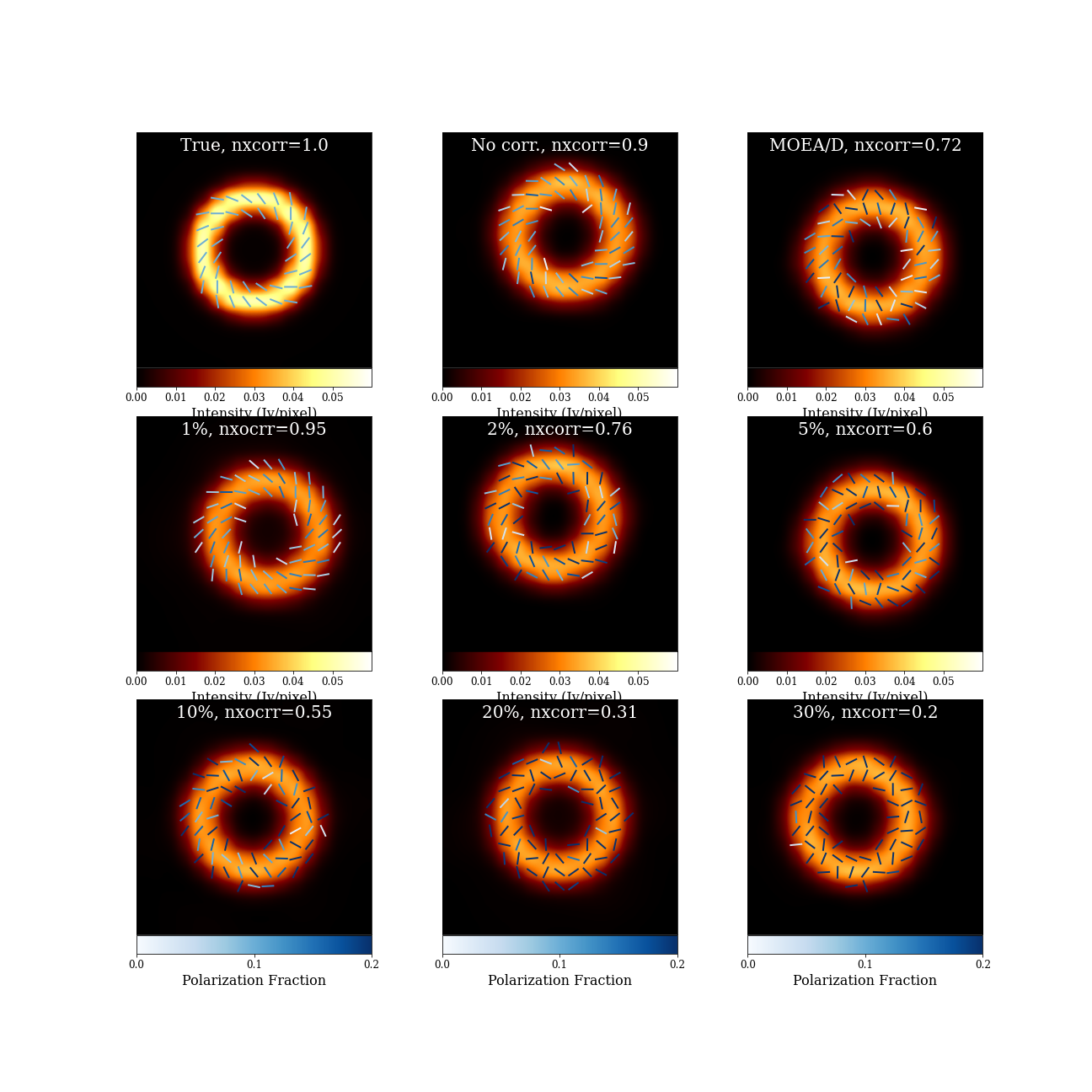}
    \caption{Comparison of the impact of residual leakage in comparison to the leakage independent imaging from the closure traces by MOEA/D. Top row: the ground truth image, the direct imaging result when leakages are perfectly recovered, and the closure trace fitting result by MOEA/D. Middle and bottom row: the direct reconstruction results with increasing level of residual leakage errors (that are not calibrated out of observation).}
    \label{fig: dterms}
\end{figure*}

Imaging directly from the closure traces has the significant advantage of being independent from any gain and leakage corruptions. However, as a disadvantage, the related optimization problem is more complex and has several potential degeneracies. In this subsection, we discuss at what point the advantages start to outweigh the disadvantages. To this end, we run a targeted test on the ring model with the EHT+ngEHT coverage. We recover the polarimetric result directly with DoG-HiT, adding no d-term corruptions, and adding $1\,\%, 2\,\%,5\,\%,10\,\%,20\%,30\%$ (residual) leakage corruptions. In reality someone would estimate the leakages based on the total intensity image or together with it. However, this reconstruction is subject to inaccuracies as well, which may lead to residual leakage errors that are passed down to the polarimetric imaging. Here we take the size of the residual leakages that were not successfully calibrated into account in comparison to the reconstruction done by closure trace imaging. To this end, we do not calibrate the leakages that we added to the data, and therefore estimate the impact of $1\,\%, 2\,\%,5\,\%,10\,\%,20\%,30\%$ \textit{residual} leakages after calibration rather than the effect of this size of leakages in general. In this way, we can explicitly quantify the effect of corrupted (or not fully calibrated) leakages on the fidelity of the image reconstruction. We evaluate the reconstructions with the cross-correlation between the ground truth and the recovered solution for $P = Q+iU$ that was proposed in \citet{eht2021a}. The results are plotted in Fig. \ref{fig: dterms}. 

As expected, the direct reconstruction with DoG-HiT with the synthetic data, that are corrupted by thermal noise and complex gains, but have the correct d-terms, performs best among all reconstructions. In the second and third column we identify the effect of dropping accuracy in the d-terms. As expected, the cross-correlation between the ground truth image and the recovered images decreases with increasing level of perturbations. The MOEA/D reconstruction becomes competitive for d-term inaccuracies of roughly $2\%$. This is indeed a typical size of inaccuracies in the determination of d-terms. If the residual d-terms after calibration are larger than $5\%$, the inaccuracy in the polarimetric calibration dominates the drop in independent observables.

\subsection{Stabilizing direct imaging}\label{sec: stabilizing}
Finally, we discuss the prospects of stabilizing the direct imaging from the Stokes visibilities with a fit to the closure traces. To this end, we are utlilizing MO-PSO, but add the $\chi^2$-statistics to the vector of objectives as a fifth objective. Now the multi-objective optimization problem consists of three regularization terms ($R_{hw}, R_{ms}, R_{ptv}$) and two data terms ($\chi^2_{CT}, \chi^2_{pvis}$). The respective reconstruction result is shown in the last column of Fig. \ref{fig: pso_ngeht} and \ref{fig: pso_eht}.

For the EHT+ngEHT configuration at which the data set is very constraining on its own only minor improvements are detected. The EHT configuration shows an interesting behaviour. The direct imaging to the non-calibrated d-terms and the simple fit to the closure traces do not produce well reconstruction. If both observables are combined however, the reconstruction performs well.

\subsection{Dependence on data quality}
In this manuscript we mainly discussed and compared two different settings, namely the setting of the EHT in 2017, and a way denser array proposed for the EHT+ngEHT. The results presented in the previous subsections suggest that directly fitting the closure traces, either with MOEA/D or MO-PSO, allows for reasonable reconstructions in the setting for the EHT+ngEHT comparable to the reconstruction with residual leakage errors of a few percent. However, as can be seen for example by comparing the reconstructions obtained with the EHT in Fig. \ref{fig: pso_eht} and the EHT+ngEHT setting in Fig. \ref{fig: pso_ngeht}, or the middle left and middle right columns in Fig. \ref{fig: comp}, the reconstruction is very challenging for the EHT configuration of 2017, and largely unsuccesful except for the case of stabilizing the fit to the polarimetric visibilities with the closure traces as discussed in Sec. \ref{sec: stabilizing}.

In this subsection, we discuss an intermediate configuration, namely an observational setting that may be expected from the 2021 EHT array. This array contains of the 2017 EHT array, and additionally the Greenland Telescope \citep[joined in 2018 and signifcantly improved the coverage in North-South direction][]{eht2024a}, the 12m telescope at Kitt Peak and NOEMA (joining in 2021). To this end, we reused the EHT+ngEHT simulation, and flagged all baselines not in the 2021 array.

\begin{figure*}
    \centering
    \includegraphics[width=\textwidth]{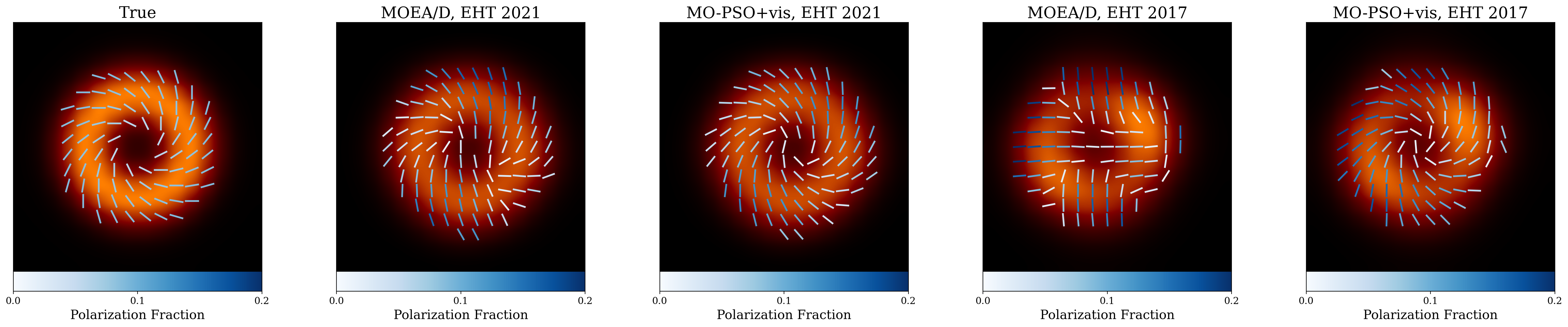}
    \caption{Ring model (let panel) recovered with MOEA/D and MO-PSO (including polarimetric visibilities) with the 2021 configuration (second and third panel). For comparison, we reprint the reconstruction results in the EHT 2017 array in the last two panels.}
    \label{fig: 2021_EHT}
\end{figure*}

We show the reconstruction results comparing to the 2017 EHT observations discussed earlier in Fig. \ref{fig: 2021_EHT}. From this short analysis, we can conclude that the direct fitting to the closure traces may already be a viable option with the current EHT array, i.e. the configuration in 2021.

\section{Real Data Application}
Most of the tests done in this work reside around synthetic EHT data. Finally, we test the procedure on existing, real observations. We apply our algorithms to the public calibrated M87 observations of the EHT at April 11th in 2017 \footnote{retrieved from \url{https://datacommons.cyverse.org/browse/iplant/home/shared/commons_repo/curated/EHTC_M87pol2017_Nov2023}}. However, we remind that it is one of our conclusions that it is challenging to emulate a consistent polarimetric image of M87 with the EHT coverage of 2017 from the closure traces alone with our procedure, compare the reconstructions shown in Fig. \ref{fig: comp} and \ref{fig: pso_eht}. Therefore, we choose a different test data set additionally. To this end, we aim to reproduce the numerical experiment performed in \citet{Albentosa2023} which represents the pioneering work that directly fitted closure traces. 

\begin{figure*}
    \centering
    \includegraphics[width=\textwidth]{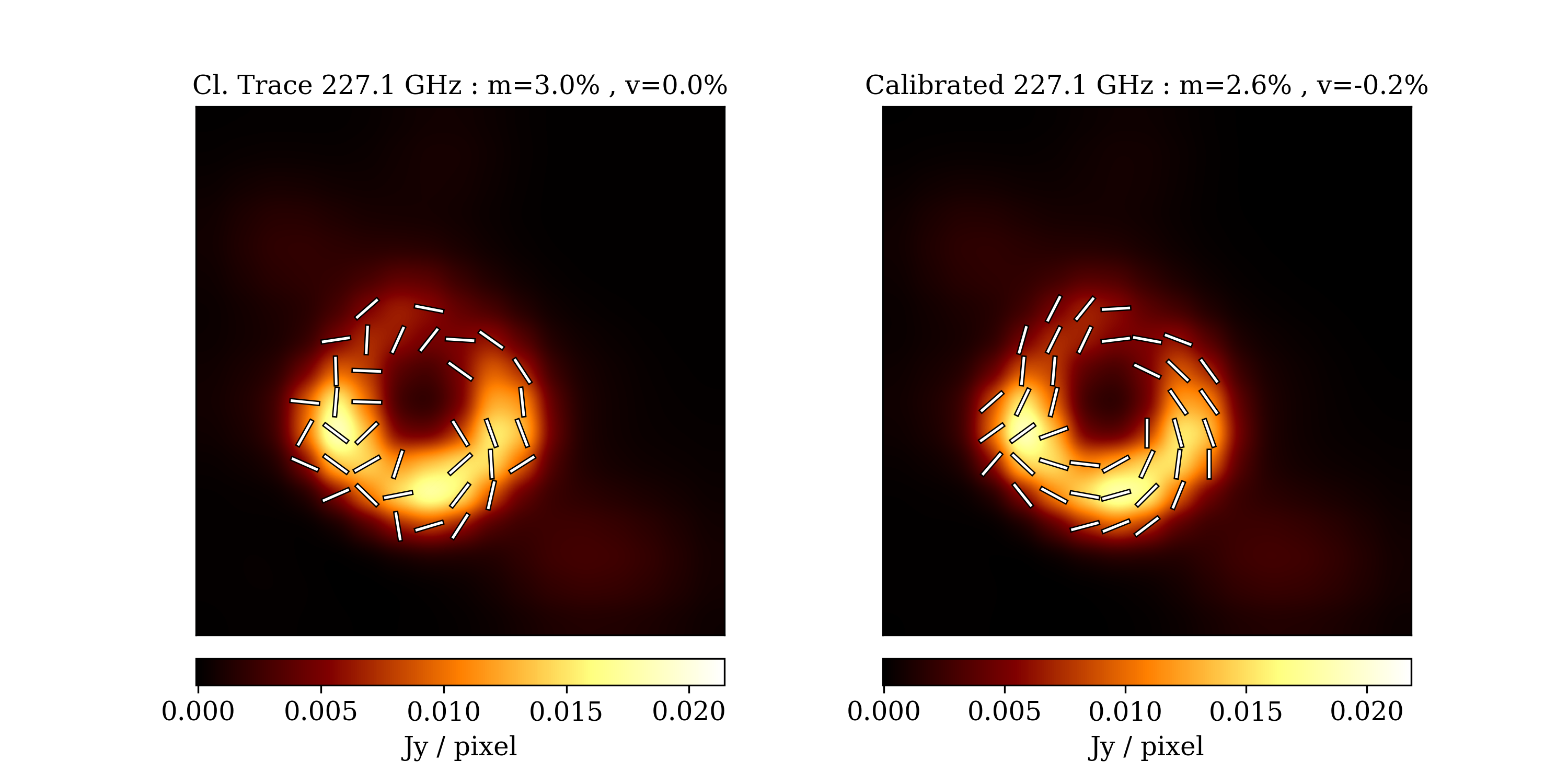}
    \caption{Reconstructions on real data observed in 2017 by the EHT. In the left panel, we show the reconstruction achieved by directly fitting closure traces with MO-PSO, on the right panel the reconstruction when fitting the Stokes visibilities with DoG-HiT, with the leakage already calibrated.}
    \label{fig: real_data}
\end{figure*}

We show in Fig. \ref{fig: real_data} the reconstruction with real data. The left panel shows the reconstruction with MO-PSO. In the right panel we show as a comparison the reconstruction done with DoG-HiT when fitting polarimetric visibilities instead. Note that in latter case the leakage terms, the main cause of issues closure trace imaging is designed to be robust against, were corrected by the detailed and vetted estimates obtained by the EHT \citep{eht2021a}. That means that the DoG-HiT reconstruction has been obtained from the already calibrated data.

Both reconstructions match reasonably well in recovering an overall azimuthal pattern, also reported by the EHT collaboration \citep{eht2021a, eht2021b}. The reconstruction done on the calibrated data however is more regular and less noisy, which may be expected by the higher signal to noise ratio in the visibility data compared to the closure traces whose information content is reduced in comparison to the visibilities.

\citet{Albentosa2023} analyze the observations of M87 that were observed with the Atacama Large Millimeter Array (ALMA) during the EHT campaign in 2017 \citep{eht2019a}, at April 6. The observations were performed in ALMA Band 6, with an array of 33 antennas covering baseline lengths between 15.1 meter and 160.7 meter. \citet{Albentosa2023} studied four spectral windows at frequencies of 213.1 GHz, 215.1 GHz, 227.1 GHz, and 229.1 GHz respectively. For this work, we reproduce the image at the third spectral window (227.1 GHz), matching the lower band of the EHT observations. 

The ALMA observations are high signal to noise reconstructions on a large number of antennas. We remind ourselves that the information loss by the use of closure trace instead of visibilities, e.g. the number of statistically independent closure traces to the number of statistically independent visibilities, converges to one with the number of available antennas \citep{Broderick2020b}. Hence, a proper reconstruction performance is expected. We like to highlight, that this application is one of the first applications of the family of closure-only imaging algorithms developed for the EHT to non-EHT data. This is particular interesting for widefield, wideband interferometric imaging at which the polarization dependence of the primary beam is taken into account by projection algorithms approximately \citep{Bhatnagar2013}. In this work, we focus on the simpler problem of reproducing the modelling results presented in \citet{Albentosa2023} instead, and leave the discussion of bypassing antenna-based primary beam pattern for a consecutive work.

The data were downloaded from the ALMA archive and calibrated with the ALMA pipeline script. We exported the measurement sets to the uvfits format and time-averaged the data by 10 seconds, and performed the reconstruction with the software tool MrBeam\footnote{\url{https://github.com/hmuellergoe/mrbeam}}. The pipeline product of ALMA was used as an initial guess. The reconstruction of the polarized structure from the closure traces, was then performed with the MO-PSO algorithm that performs much faster than MOEA/D. Due to the larger number of pixels and the larger number of visibilities, this speed-up proved to be essential to ensure timely numerical performance.

\begin{figure*}
    \centering
    \includegraphics[width=0.6\textwidth]{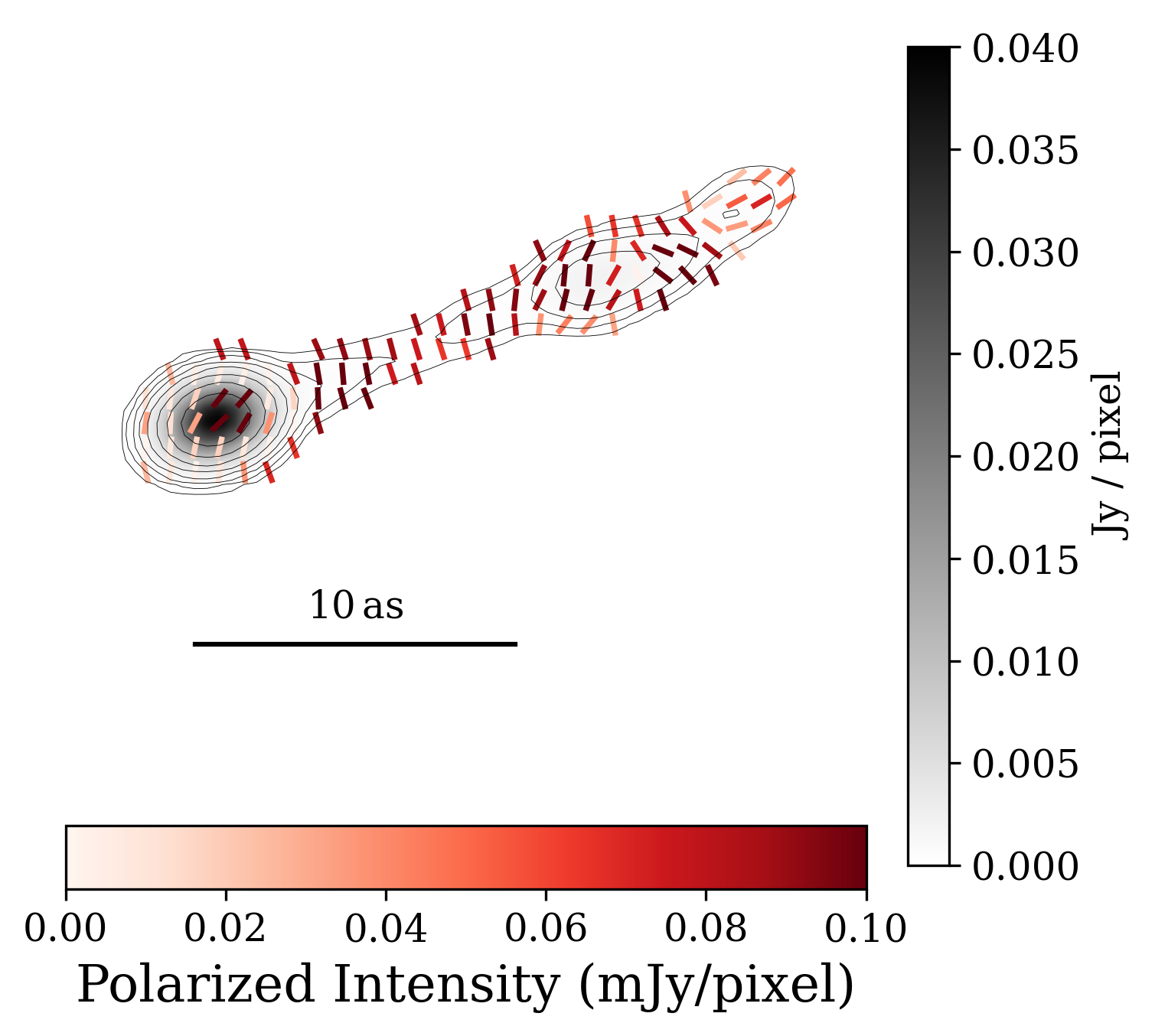}
    \caption{Reconstruction of ALMA Band 6 observations of M87 at 227.1 GHz.}
    \label{fig: alma}
\end{figure*}

We show the recovered polarization structure in Fig. \ref{fig: alma}. The Stokes I reconstruction, as well as the EVPA structure looks very similar to the reconstructions presented in \citet{Albentosa2023}, presenting an interesting cross-verification. We would like to note, that the approach presented in this manuscript differs significantly from the approach pioneered in \citet{Albentosa2023}. Latter one used a model-fitting approach and explored the model-fitting components with a full Markov chain Monte Carlo (MCMC) approach, while we study the direct imaging with multiobjective optimization. Given these philosophically differences in the way the problem is approached, the significant similiarities in the reconstructions serve as a powerful consistency check of the procedure described in this manuscript. Finally, we would like to note that \citet{Albentosa2023} found multiple modes describing the data when performing a full gridded exploration of the $\chi^2$-parametric space. While we do not reproduce the multimodality of the problem setting since we are using MO-PSO instead of MOEA/D (which would explore the full Pareto front), MO-PSO seems successful in navigating this challenging parametric space due to its evolutionary approach.

The reconstruction was performed on a common notebook in roughly 30 minutes. This highlights the scalability of the MO-PSO algorithm to existing data sets, and the applicability of procedures that were initially proposed for the EHT in a more general setting.

\section{Conclusions}
Closure traces \citep{Broderick2020b} are a concept extending the well known closure phases and closure amplitudes to the polarimetric domain. As a key feature, it can be shown that they are independent against gain corruptions and leakages, marking them an important closure quantity to study the polarimetric signature of a source. This is particularly true for the EHT and its planned successors (e.g. the EHT+ngEHT) due to the relative sparsity of the observation, the low signal-to-noise ratio and the huge gain uncertainties. In this manuscript, we evaluate with some exemplary studies whether closure traces could be used as a data functional in the same way closure phases and closure amplitudes are utilized for closure-only imaging. The importance of this operational mode may stem particularly from the potential lack of suitable calibrators with sufficient parallactic angle coverage, or the lack of unresolved calibrator sources at all. A first pioneering work in this direction has been presented by \citet{Albentosa2023}, albeit in a model-fitting setting rather than by direct imaging. Here, we study the imaging problem instead.

To this end, we construct an imaging strategy that explicitly deals with the degeneracies and potential multimodality that are associated with closure traces. In a nutshell, we utilize the multiobjective imaging methods MOEA/D \citep{Mueller2023c, Mus2023b} and MO-PSO \citep{Mus2023c} that approximate multiple local minima simultaneously. In fact, multiple local minima were identified (among the rotation invariance around the Poincare-sphere), that would pose significant challenges for local search techniques.

We applied our imaging pipeline to three different synthetic data sets resembling usual EHT targets, in two different array configurations and with various levels of noise corruptions. As a remaining limitation, we rely on the assumption of vanishing circular polarization which may not be justified for the main targest of the EHT \citep{eht2023}.

This application demonstrated that the closure traces, while putting reliable upper limits on the polarization fraction \citep{eht2021a, eht2023}, are not informative enough for the EHT to allow for polarimetric imaging. This is consistent with the conclusions drawn by the EHT collaboration \citep{eht2021a}, mainly caused by the small number of quadrilaterals that probe extended structural information with sufficient signal to noise ratio.

A planned successor of the EHT, the EHT+ngEHT with ten additional antennas, however has a much larger number of quadrilaterals. Closure trace imaging becomes a feasible alternative in this configuration, and indeed produces results that are competitive to a few percent of residual leakages. Similarly, already the EHT configuration of 2021 may produce a statistical set large enough to apply this technique.

Although the data are more constraining with the additional antennas assumed in this work, the Pareto front recovered by MOEA/D still features multiple prominent clusters. This can be interpreted by the existence of multiple local minima, that (depending on the starting point) a local search technique may fall into. The cluster that is closest to the utopic ideal \citep[as proposed by][]{Mueller2023c, Mus2023c} contains the best solution. This supports the use of MO-PSO, which is a faster alternative to MOEA/D, that utilizes the distance to the ideal as its objective functional. 

Complementing these tests on synthetic data in an EHT setting, we apply the pipeline to real observations with ALMA on M87. The recovered structure is similar to the model-fitting approach presented in \citet{Albentosa2023}, further substantiating the robustness of the data analysis pipeline from the closure traces alone presented in this manuscript. Moreover, this application demonstrates the scalability of MO-PSO to less sparse, high-dynamic range data sets, and contributes to a line of recently started research evaluating the usefulness data analysis techniques proposed by the EHT in a more general setting.

We conclude that closure trace imaging will be a feasible and competitive alternative to the classical hierarchical VLBI imaging, or the simultaneous reconstruction of gains and d-terms together with the image structure are explored by Bayesian methods, at least in the numerical practice when theoretical marginalization over the unknown gains and leakage terms may include non-trivial approximations. In this manuscript, we pointed out the strategy how to implement this philosophy in practice. Finally, closure trace imaging does not need to be applied independently, a short analysis showed that it may also improve the RML imaging pipeline by being applied together with the classical polarization data terms. 

\begin{acknowledgement}
This work was partially supported by the M2FINDERS project funded by the European Research Council (ERC) under the European Union’s Horizon 2020 Research and Innovation Programme (Grant Agreement No. 101018682). HM received financial support for this research from the International Max Planck Research School (IMPRS) for Astronomy and Astrophysics at the Universities of Bonn and Cologne, and through the Jansky fellowship awarded by the National Radio Astronomy Observatory. The National Radio Astronomy Observatory is a facility of the National Science Foundation operated under cooperative agreement by Associated Universities, Inc. This paper makes use of the following ALMA data: ADS/JAO.ALMA$\#$2016.1.01154.V ALMA is a partnership of ESO (representing its member states), NSF (USA) and NINS (Japan), together with NRC (Canada), MOST and ASIAA (Taiwan), and KASI (Republic of Korea), in cooperation with the Republic of Chile. The Joint ALMA Observatory is operated by ESO, AUI/NRAO and NAOJ. The data underlying this work are accessible from the ALMA data archive, synthetic data sets will be shared upon reasonable request. The software used to produce these results is MrBeam, publicly available through github\footnote{\url{https://github.com/hmuellergoe/mrbeam}}. MrBeam makes explicit use of the following packages: ehtim\footnote{{https://achael.github.io/eht-imaging/}}, WISE\footnote{\url{https://github.com/flomertens/wise}}, regpy\footnote{\url{https://num.math.uni-goettingen.de/regpy/}} and pygmo\footnote{\url{https://esa.github.io/pygmo2/}}.

\end{acknowledgement}

\bibliography{lib}{}
\bibliographystyle{aa}

\appendix

\section{Description of the imaging algorithms}\label{app:imaging}
\subsection{MOEA/D}
RML imaging algorithms minimize a sum of data fidelity terms and regularization terms. By minimizing a weighted sum of these terms, one tries to find a solution that is favorable both by the data fidelity terms and the regularization terms, i.e. that fits to the observed data and is physically reasonable. There are multiple regularization terms that are commonly used for VLBI representing a wide variety of prior assumption, e.g. simplicity of the solution (measured for example by the entropy $R_{\text{entr}}$ or the $l^2$ norm $R_{l^2}$), sparsity ($l^1$ norm $R_{l^1}$), smoothness (total squared variation $R_{\text{tsv}}$, total variation $R_{\text{tv}}$) or the correct total flux. We refer to \citet{Mueller2023c} and \citet{eht2019d} for a discussion of the full range of regularization terms.

For MOEA/D, the whole range of reasonable image features that may be fitted to the data with a specific regularization parameter weight combination is searched by multiobjective evolution where every data term and regularization term is modelled as an individual objective \citep{Mueller2023c}.

MOEA/D recovers the Pareto front consisting of all Pareto optimal solutions by the genetic algorithm \citep{Zhang2008, Li2009}. A solution is called Pareto optimal if the further optimization along one of the objectives has to worsen the scoring in another objective, e.g. the further minimization of the $\chi^2$ metric has to worsen the entropy or sparsity assumption. In this way, MOEA/D succeeds to explore the image structures globally, recovering all the locally optimal image structures along the objectively best.

It has been shown in \citet{Mueller2023c} that the Pareto front has a specific structure for an EHT-like configuration: it divides into multiple disjoint clusters, each of them representing a locally optimal mode. 
For more details we refer to \citet{Mueller2023c} and \citet{Mus2023b} and references therein. 


\subsection{MO-PSO}
\begin{figure}
    \centering
    \includegraphics[width=0.5\textwidth]{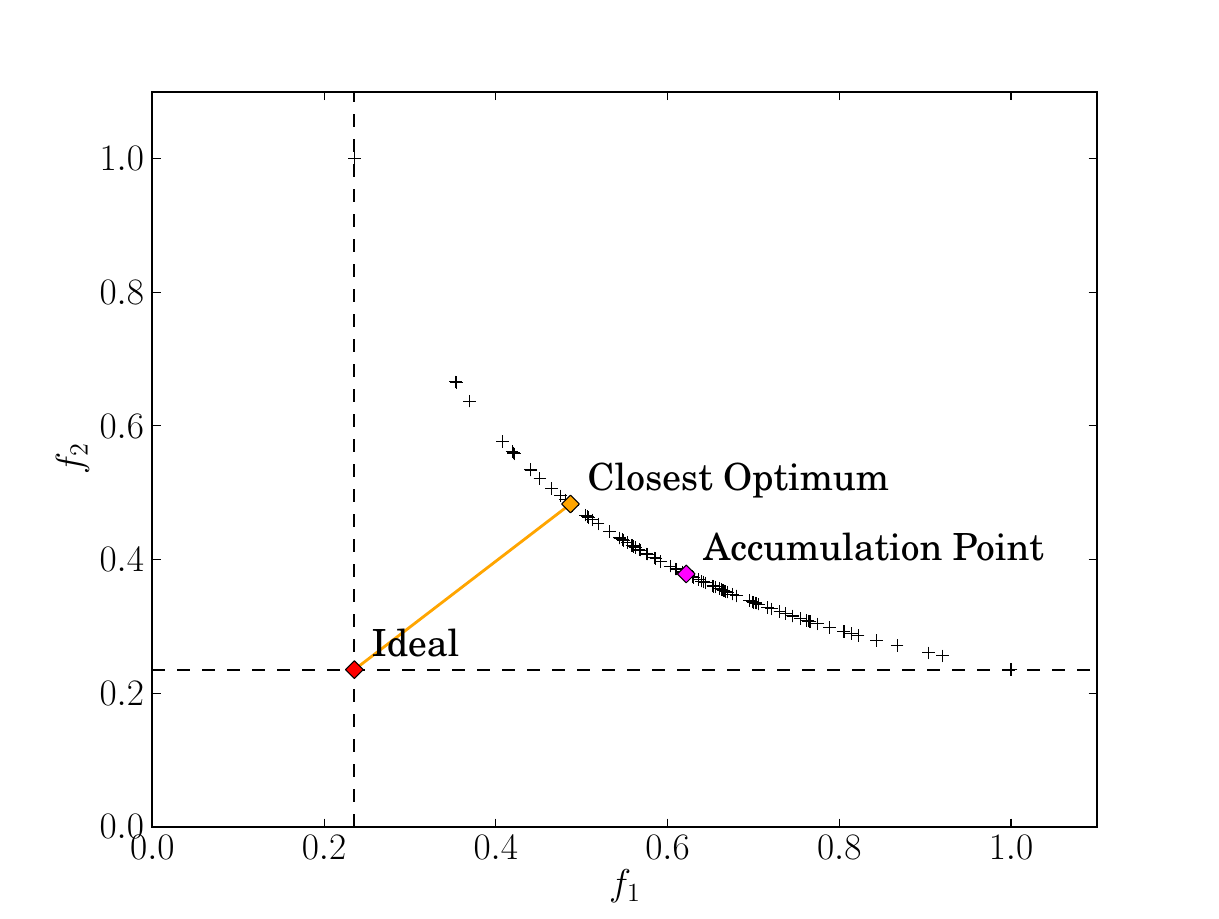}
    \caption{Illustrative sketch of the Pareto front in two dimensions, the construction of the ideal point by the asymptotic behaviour among every of the single axes, and the distance to the Pareto front as a selection criterion. This distance is minimized by MO-PSO. The figure is reprinted from \citet{Mueller2023c} under CC.BY. 4.0..}
    \label{fig:pareto-scheme}
\end{figure}

MO-PSO was proposed by \citet{Mus2023c} as a companion to MOEA/D. It solves some of the numerical limitations of MOEA/D and improves the accuracy of the reconstruction. The Pareto front has an asymptotic behaviour with regard to any axis, i.e. individual objective. This is the scoring of any of the objective terms when only this objective plays a role in the reconstruction. 
The stacked asymptotic scoring in any of the objective terms is the ideal point (see Fig. \ref{fig:pareto-scheme} for an illustrative sketch). 
It has been shown in \citet{Mueller2023c} and \citet{Mus2023c} that the distance between the ideal point and the scoring of the Pareto optimal solutions can be used to select the best reconstruction in the Pareto front. 

MO-PSO uses this observation to define a convex optimization problem with respect to the weights. It tries to find the Pareto optimal image structure with the smallest distance to the ideal point (i.e. the closest optimum in Fig. \ref{fig:pareto-scheme}) by screening the regularization weights with a particle-swarm optimization algorithm, and solving the related weighted sum problem by fast gradient based minimization techniques. For more details, we refer to \citet{Mus2023c}. In a nutshell, every particle in the swarm represents a single weight combination. We solve the related weighted sum problem. 
This solution is approximating a Pareto optimal solution. Next we compute the distance of this solution to the ideal point. Then we update the position of every particle, i.e. the weight combinations, by the update step of a particle swarm optimization, solve for the related images, and accept these updates if the distance to the ideal point has improved.

\subsection{DoG-HiT}
DoG-HiT models the image structure by wavelets that are fitted to the uv-coverage \citep{Mueller2022}.
The image is recovered by standard sparsity promoting regularization.
In this way, DoG-HiT effectively computes the multiresolution support, that is the set of statistically significant wavelet coefficient that are needed to represent the image, inspired by \citet{Mertens2015}. 
It has been demonstrated in \citet{Mueller2023b} that the multiresolution support is a reasonable prior for a full polarimetric (and time-dynamic) reconstruction. For the reconstruction of the polarimetric signal, we only vary the wavelet coefficients within the multiresolution support and fit the Stokes visibilities directly, i.e. we solve the problem by a constrained minimization approach.

DoG-HiT is not the primary subject of the investigation in this manuscript. In this manuscript, we fit directly to the closure traces with MOEA/D and MO-PSO. However, we use DoG-HiT as an initial guess for the total intensity structure, and as a comparison, imaging algorithm. DoG-HiT is a good choice for a comparison of the performance, since it is philosophically close to the multiobjective approaches (i.e. it approaches the problem in a RML fashion, rather than by inverse modelling or Bayesian exploration), it is fast, unbiased in total intensity (fits only to closure phases and closure amplitudes), and, most importantly, it is automatized and unsupervised \citep[e.g. see the discussion in][]{Mueller2022}. Particularly the last point makes it a reasonable choice as a comparison method, aiming at minimizing the human factor in the comparison. Moreover, the reconstruction of the linear polarization is straightforward and represents a relatively weak, but data-driven, prior assumption, opposed to the computationally expensive algorithms MOEA/D and MO-PSO that survey a variety of strong prior assumptions. 

\section{Polarimetric Regularization Terms}\label{app:regularizers}
The polarimetric regularizers used in this manuscript are as follows: We use polarimetric variants of the image entropy as regularizers, as proposed by \citet{Narayan1986}:
\begin{align} \nonumber
    &R_{hw} := \displaystyle{\sum^{N^2}_{i}} I_i \left( \ln(\dfrac{I_i}{M_i}) + \dfrac{1+m_i}{2} \ln\left(\dfrac{1+m_i}{2}\right) \right.\\
    &\hspace{2cm}\left. + \dfrac{1-m_i}{2} \ln\left(\dfrac{1-m_i}{2}\right)\right), \label{eq:shw} \\
    &R_{ms} = \displaystyle{\sum^{N}_{i}}\lvert I_i\rvert \ln \lvert m_i \rvert. \label{eq:msimple}
\end{align} 
Moreover, we are using the polarimetric counterpart of the total variation regularizer \citep{Chael2016}:
\begin{equation}\label{eq:tv}
    R_{ptv} = \sum_{i}\sum_{j}\sqrt{\vert P_{i+1, j} - P_{i,j} \rvert^2 + \left| P_{i, j+1} - P_{i,j} \right|^2}.
\end{equation}
Here $I$ denotes the total intensity, $M$ the total intensity of a prior model, $m$ the polarization fraction and $P = Q+iU$ the complex-valued linear polarization vector.

\section{Updates to the MOEA/D pipeline}\label{app:moead_updates}
In the following list we enumerate the detailed changes that needed to be done to the MOEA/D pipeline to allow for the reconstruction of the linear polarization signal directly from the closure traces:
\begin{itemize}
    \item In this manuscript, we study the prospects of using closure traces directly to fit in a self-calibration independent way the linear polarization signature, ignoring circular polarization for now, and not focusing on total intensity. Closure traces contain classical closure amplitudes and closure phases as a special case, and extend these concepts to the polarimetric domain. Instead of using closure traces only for the polarimetric signal, one could also try to recover the total intensity image by directly fitting closure traces. However, the total intensity imaging is only weakly affected by the leakages, and the imaging in total intensities by closure properties is a intensively studied problem, that leads to quite reliable image reconstructions already \citep[e.g.][]{Chael2018, Mueller2022, Arras2022, Mueller2023c, Thyagarajan2023}. Hence, we focus in this manuscript on the reconstruction of the linear polarization features by fitting the closure traces, and recover the total intensity image by alternative closure-only imaging algorithms. For the test-cases presented here, we did the reconstruction of the total intensity image by DoG-HiT fitting directly closure amplitudes and closure phases.
    \item Once the total intensity image has been successfully recovered, we move to the polarimetric part. The polarimetric signal is represented by the polarization fraction $m$ and the mixing angle $\chi$ for the polarimetric reconstruction rather than by $Q$ and $U$. This equivalent reparametrization allows to utilize the inequality \eqref{eq: pol_ieq} in a more straightforward way by rigidly setting bounds on the polarization fraction.
    \item Originally, MOEA/D is either initialized by a completely random initial population, or by the result of an unpenalized imaging run with few iterations. \citet{Mueller2023c} provides a comparison between these two initialization strategies. The first one starts with a more global distribution of genomes across the parameter space, making it easy to recover the potentially multimodality of the problem, latter strategy speeds up the reconstruction with MOEA/D with a starting point that is actually closer to the finally preferred solution. In this manuscript, we opted for a hybrid way. Since the unpenalized minimization varies significantly with the initial guess used, we used 11 random EVPA orientations, performed the unpenalized minimization, and initialized the initial population for MOEA/D by 91 copies of each, matching the overall population size 1001 genomes in the populations that is set by the combinatoric number of weightings $\lambda$ with a fixed grid size that sum to one.
    \item To avoid the situation that due to this random initialization, specific source solutions are related to specific weight combinations, we added a fifth functional (i.e. a fifth axis) to MOEA/D. This functional is constantly zero, so does not add anything up to the optimization at all, but serves as an axis to store multiple solution varieties originating from various starting points for the same weighting combination. Furthermore, we randomize the weight generation $\lambda$ rather than drawing them from a grid as we have done before.
    \item Since closure traces are independent against any global rotation of the EVPA pattern, MOEA/D naturally samples the full domain of the global orientations, i.e. the population contains genomes that only differ by an overall constant offset to the mixing angle $\chi$. All these images that are equal up to a constant angle, have the same scoring in the objective functionals since also the regularizer terms are independent of the overall orientation. To avoid this unnecessary duplication of information we are using an alternating procedure. We take the initial population, update the population by a number of generations (set to 10), order the images in clusters as described in \citet{Mus2023b}, and align all genomes within one cluster with respect to the overall EVPA orientation. Then we initialize the next population with the aligned genomes, and proceed the genetic evolution. This procedure is repeated multiple times.
\end{itemize}

\section{Pareto fronts} \label{app:pareto_fronts}

\begin{figure*}
    \centering
    \includegraphics[width=\textwidth]{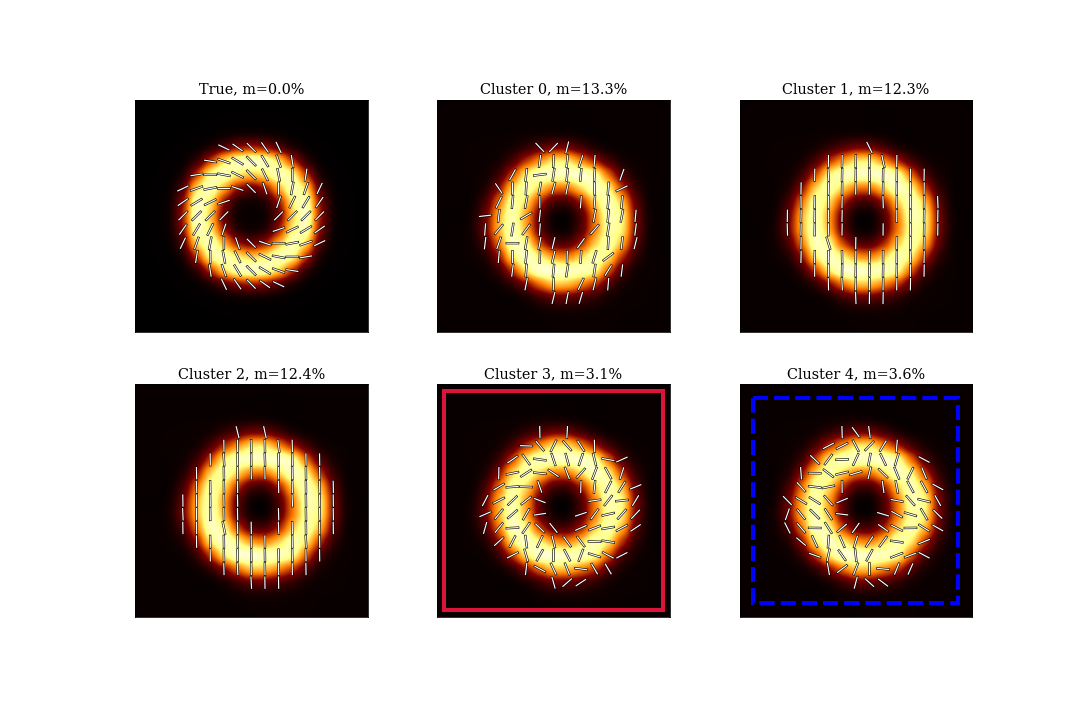}
    \caption{Pareto front of the MOEA/D reconstruction directly fitting to closure traces for the EHT+ngEHT array configuration. The solution in the red box is preferred by MOEA/D as the Pareto-optimal solution closest to the optimum, the cluster with the blue box would have been selected when comparing to closure quantities defined in \citet{Thyagarajan2022, Joseph2022}.}
    \label{fig: model1}
\end{figure*}

\begin{figure*}
    \centering
    \includegraphics[width=\textwidth]{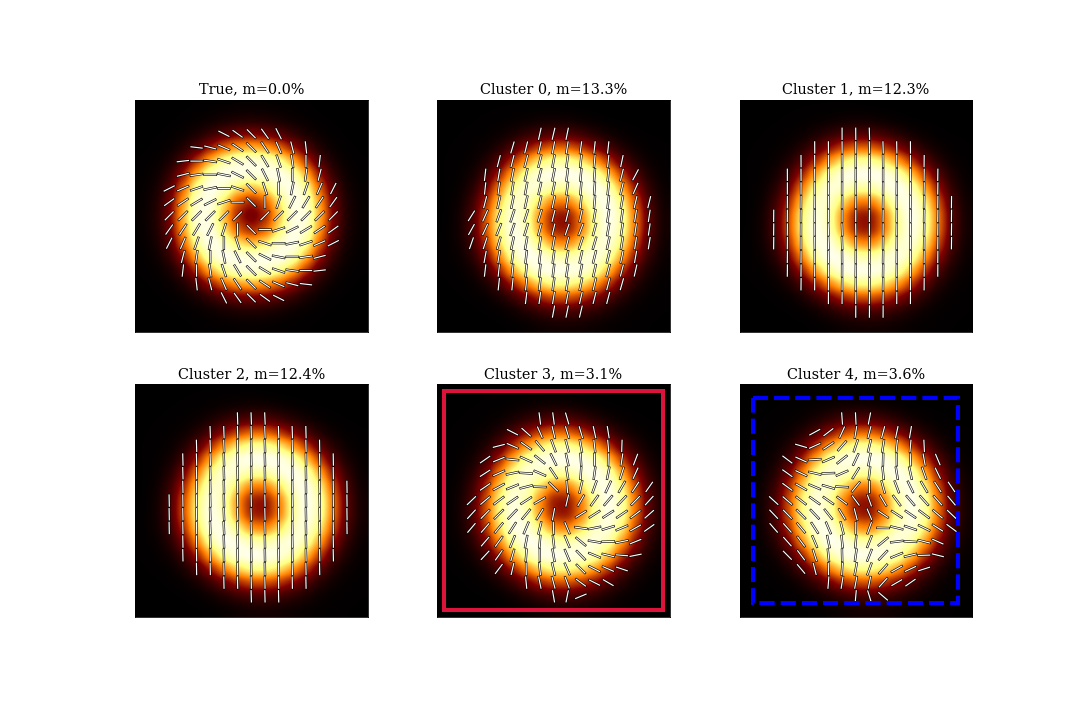}
    \caption{Fig. \ref{fig: model1}, but with a blurring of $20\,\mu\mathrm{as}$ applied to every cluster.}
    \label{fig: model1_blurred}
\end{figure*}

\begin{figure*}
    \centering
    \includegraphics[width=\textwidth]{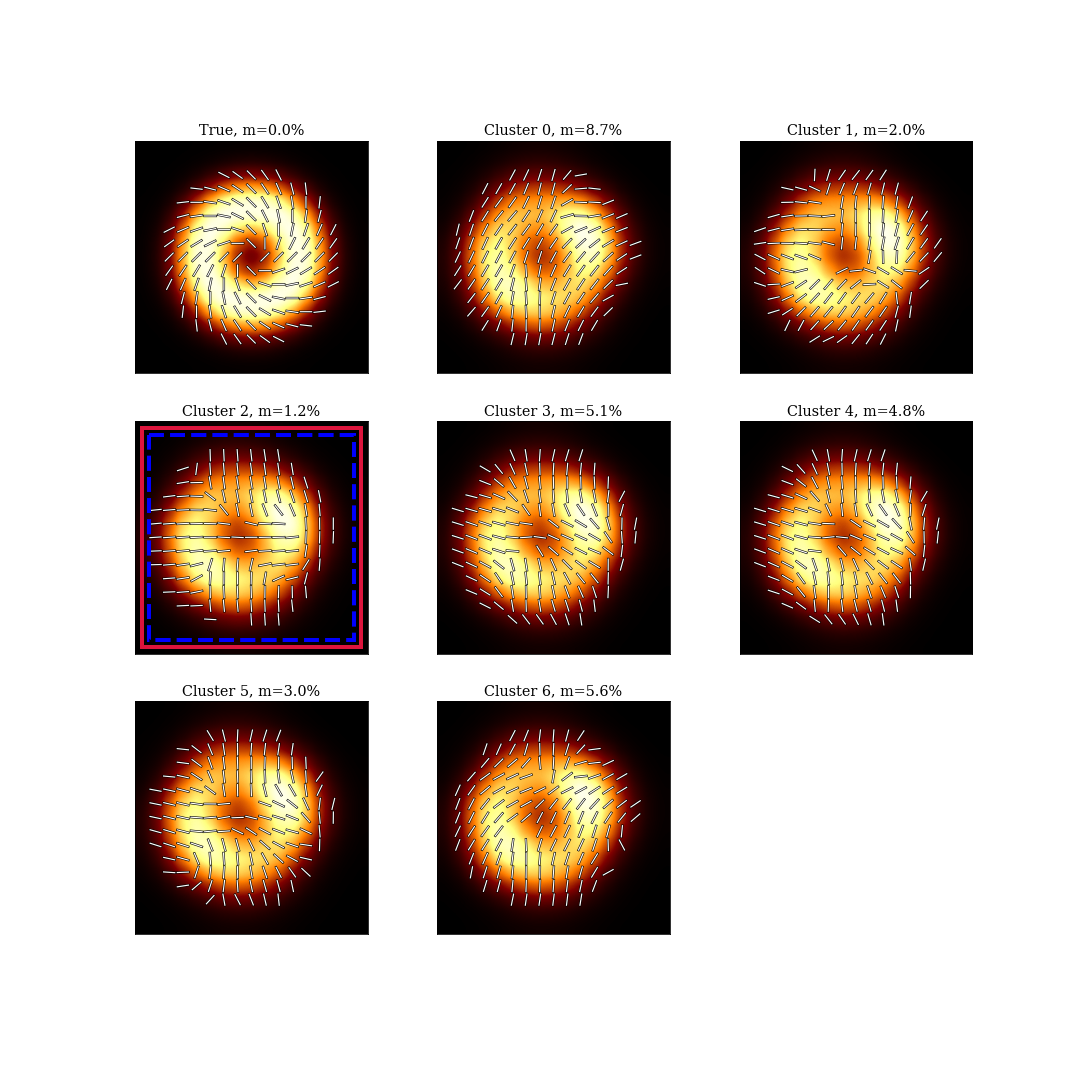}
    \caption{The same as Fig. \ref{fig: model1}, but for the EHT array configuration. Additionally a blurring of $20\,\mu\mathrm{as}$ has been applied to every cluster.}
    \label{fig: model1_eht_blurred}
\end{figure*}

\begin{figure*}
    \centering
    \includegraphics[width=\textwidth]{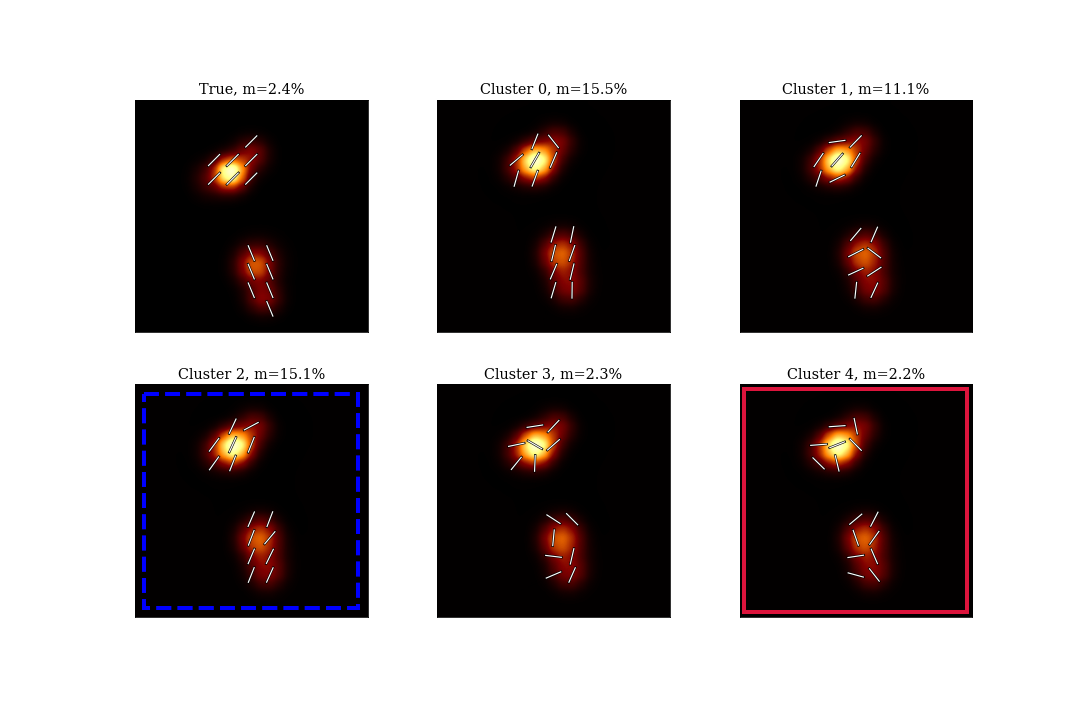}
    \caption{The Pareto front of the double source in the EHT+ngEHT configuration.}
    \label{fig: double}
\end{figure*}

\begin{figure*}
    \centering
    \includegraphics[width=\textwidth]{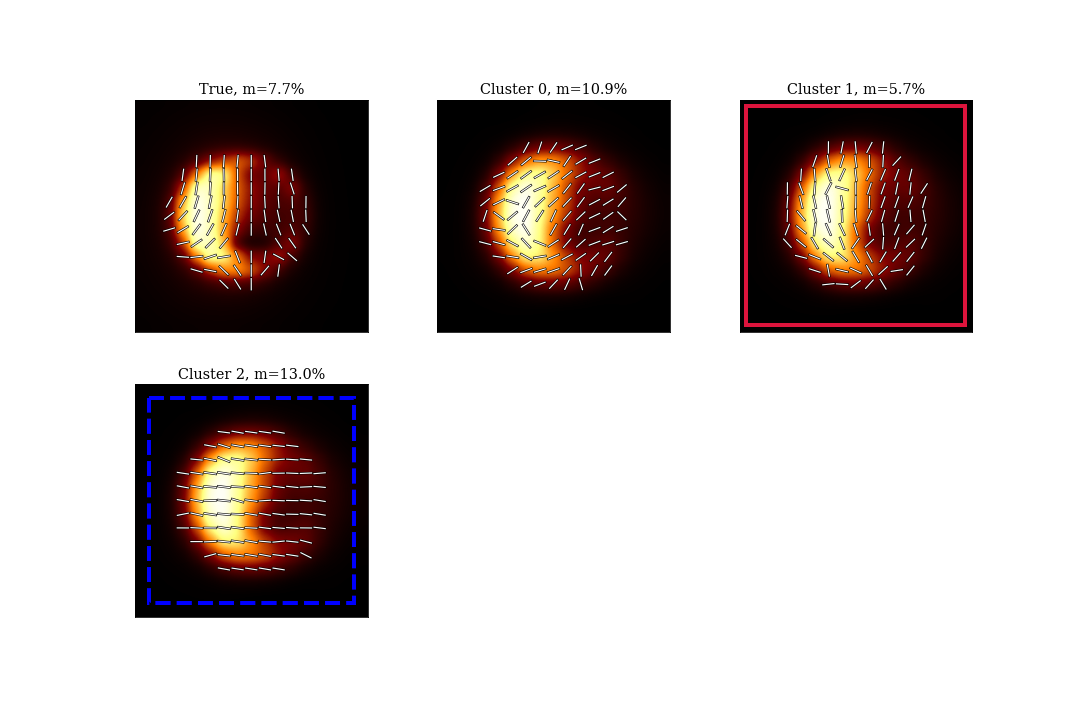}
    \caption{The Pareto front of the SgrA* source model in the EHT+ngEHT configuration.}
    \label{fig: sgra}
\end{figure*}

\end{document}